\documentclass[fleqn,usenatbib]{mnras}

\usepackage{newtxtext,newtxmath}

\usepackage[T1]{fontenc}

\DeclareRobustCommand{\VAN}[3]{#2}
\let\VANthebibliography\thebibliography
\def\thebibliography{\DeclareRobustCommand{\VAN}[3]{##3}\VANthebibliography}

\newcommand{\Msun}{M_\odot}
\newcommand{\Rsun}{R_\odot}
\newcommand{\Mbh}{M_\bullet}
\newcommand{\Mdot}{\dot M}
\newcommand{\Medd}{\dot M_{\rm Edd}}
\newcommand{\tfb}{t_{\rm fb}}
\newcommand{\torb}{t_{\rm orb}}
\newcommand{\tvisc}{t_{\rm visc}}
\newcommand{\Rcirc}{R_{\rm circ}}

\newcommand{\rT}{R_{\rm T}}
\newcommand{\rg}{r_g}
\newcommand{\Mdotfb}{\Mdot_{\rm fb}}
\newcommand{\Mdotacc}{\Mdot_{\rm acc}}

\usepackage{graphicx}	
\usepackage{amsmath}	
\usepackage{newtxtext,newtxmath}
\usepackage{array}
\usepackage{makecell}
\usepackage{longtable}
\usepackage{adjustbox}

\title[Radio observations of GRB250702B]{One year of broadband radio monitoring of the enigmatic transient GRB 250702B reveals the evolution of the relativistic jet}

\author[A. J. Goodwin et al.]{ A. J. Goodwin,$^{1}$\thanks{E-mail: ajgoodwin.astro@gmail.com} J. C. A. Miller-Jones,$^{1}$ Itai Sfaradi,$^{2,3}$ A. Mummery,$^{4}$ Raffaella Margutti,$^{2,3,5}$ Tanmoy Laskar,$^{6}$
\newauthor{K. D. Alexander,$^{7}$ Yuhan Yao,$^{8}$ Arvind Balasubramanian,$^{9}$ G. C. Anupama,$^{9}$ Edo Berger,$^{10}$ Varun Bhalerao,$^{11}$ }
\newauthor{Yvette Cendes,$^{12,13}$ Ryan Chornock,$^{2,3}$ C. T. Christy,$^{7}$ D. Eappachen,$^{6,9}$ Tarraneh Eftekhari,$^{14}$} 
\newauthor{Miguel Pérez-Torres,$^{15}$ Enrico Ramirez-Ruiz,$^{16}$ D. K. Sahu,$^{9}$ and Sjoert van Velzen$^{17}$}\\ 
$^{1}$International Centre for Radio Astronomy Research -- Curtin University, GPO Box U1987, Perth, WA 6845, Australia\\ $^{2}$Department of Astronomy, University of California, Berkeley, CA 94720-3411, USA\\ $^{3}$Berkeley Center for Multi-messenger Research on Astrophysical Transients and Outreach (Multi-RAPTOR), University of California, Berkeley, CA 94720-3411, USA\\ $^{4}$School of Natural Sciences, Institute for Advanced Study, 1 Einstein Drive, Princeton, NJ 08540, USA\\ $^{5}$Department of Physics, University of California, 366 Physics North MC 7300, Berkeley, CA 94720, USA\\ $^{6}$Department of Physics \& Astronomy, University of Utah, Salt Lake City, UT 84112, USA\\ $^{7}$Department of Astronomy/Steward Observatory, 933 North Cherry Avenue, Rm. N204, Tucson, AZ 85721-0065, USA\\ $^{8}$Miller Institute for Basic Research in Science, 206B Stanley Hall, Berkeley, CA 94720, USA\\ $^{9}$Indian Institute of Astrophysics, II Block Koramangala, Bengaluru 560034, India\\ $^{10}$Center for Astrophysics $|$ Harvard \& Smithsonian, 60 Garden St., Cambridge, MA 02138, USA\\ $^{11}$Department of Physics, Indian Institute of Technology Bombay, Powai, Mumbai 400076, India\\ $^{12}$Department of Physics and Astronomy, University of Oregon, 1371 E 13th Ave, Eugene OR 97403, USA\\ $^{13}$Institute for Fundamental Science, University of Oregon, 1371 E 13th Ave, Eugene OR 97403, USA\\ $^{14}$Center for Interdisciplinary Exploration and Research in Astrophysics (CIERA), Northwestern University, 1800 Sherman Avenue, Evanston, IL 60201, USA\\ $^{15}$Instituto de Astrofísica de Andalucía (IAA-CSIC), Glorieta de la Astronomía s/n, E-18008 Granada, Spain\\ $^{16}$Department of Astronomy and Astrophysics, University of California, Santa Cruz, CA 95064, USA\\ $^{17}$Leiden Observatory, Leiden University, P.O. Box 9513, NL-2300 RA Leiden, The Netherlands }

\date{Accepted 2025 November 5. Received 2026 November 4; in original form 2026 August 4}

\pubyear{\the\year{}}

\begin{document}
\label{firstpage}
\pagerange{\pageref{firstpage}--\pageref{lastpage}}
\maketitle

\begin{abstract}
We present an extensive radio monitoring campaign of the unique extragalactic transient GRB 250702B, associated with unprecedentedly long duration gamma-ray emission, lasting over 3\,hr. Our observations span 0.65--233\,GHz from 6--356\,d (observer frame) post-discovery. The radio emission shows a smoothly evolving peaked synchrotron spectrum consistent with an adiabatic shock expanding into a stratified ambient medium ($n_e\propto R^{-k}$; $k= 1.5-2$). We detect significant variability in the low frequency ($\leq3$\,GHz) light curves which we interpret as interstellar scintillation, placing an approximate bound on the blast wave image size of $1.2\times10^{16}\lesssim R_{\perp} \lesssim 5\times10^{17}$\,cm. 
The temporal evolution of the flux density and critical synchrotron frequencies suggest the shock that powers the radio emission is potentially a wide-angle $\theta_j\gtrsim15$\,deg, low Lorentz factor ($\Gamma\lesssim10$) jet, or a narrow $\theta_j\lesssim2$\,deg ($\Gamma_0>50$) highly relativistic jet. A narrow jet is expected for a stellar-mass black hole engine, such as a helium star merger, and the beaming-corrected kinetic energy in this scenario is consistent with the known distribution for long GRBs ($E_K\sim10^{51}$\,erg).
The wide-angle jet scenario would instead require a progenitor involving prolonged accretion. We derive and show an intermediate or stellar-mass black hole tidal disruption event are viable possibilities. The beaming-corrected kinetic energy in this scenario is on the low end of the known distribution for relativistic SMBH TDEs ($E_K\sim10^{50}$\,erg). We disfavour an SMBH TDE due to lack of compatibility with the observed timescales. The detection of a jet shut off within the next year would favour a WD-IMBH TDE due to the shorter theoretical duration of super-Eddington accretion than the main-sequence TDE channels.
\end{abstract}

\section{Introduction}

Gamma-ray bursts (GRBs) are energetic pulses of gamma-rays typically lasting tens of seconds and observed at cosmological distances \citep{Meszaros2006,Kumar2015}. Long GRBs (lGRBs; prompt gamma-ray durations $\gtrsim2$\,s) are thought to arise during the collapse of massive stars and the resulting rapid accretion by the nascent compact object of stellar material \citep{Woosley1993,Wijers1997,MacFadyen1999}. The duration of the prompt GRB emission is presumably related to the lifetime of the central engine following jet breakout from the progenitor star \citep[e.g.,][]{Bromberg2012}. Long-lived multi-wavelength afterglow emission is commonly detected after a GRB, consistent with synchrotron radiation from a relativistic jet interacting with the circumburst medium \citep[e.g.,][]{Sari1998,Rhoads1999}.

A sub-set of GRBs, called ultra-long GRBs (ulGRBs), show prompt GRB emission lasting thousands to tens of thousands of seconds, requiring extended central-engine activity and large energy budgets \citep{Virgili2013,Stratta2013,Gendre2013,Greiner2015}. The exceptional properties of some ulGRBs have prompted alternative progenitors to be suggested for these events, including exotic supernovae \citep{Nakauchi2013}, a rare helium star merger \citep{Neights2026}, a neutron star stellar core merger \citep{HutchinsonSmith2024}, or relativistic jetted tidal disruption events (TDEs), in which an on-axis relativistic jet is launched when a star is destroyed by a massive black hole \citep{Stone2019} or a stellar mass black hole \citep[``micro-TDE";][]{Perets2016}. However, GRB 220627A, a recent high redshift ($z=3.08$) ulGRB, was found to be observationally consistent with the normal long GRB population \citep{dewet2023,Leung2026}. The TDE scenario for some hard X-ray-emitting transients was confirmed with the relativistic jetted TDE Sw J1644+57 \citep{Burrows2011,Bloom2011,Levan2011,Zauderer2011,Berger2012}, although not an ulGRB. This event initially triggered \textit{Swift} BAT as GRB 110328A \citep{Burrows2011}. It was later classified as a TDE based on its precise localisation to the nucleus of its host galaxy \citep{Zauderer2011}, long-lived and highly variable X-ray emission \citep{Levan2011}, a characteristic $t^{-5/3}$ decay \citep{Burrows2011} consistent with fallback accretion \citep[although many TDEs do not show a decay exactly consistent with the fallback rate, e.g.][]{Guolo2024,Yao2023}, and the detection of a newly launched relativistic jet which shut off $\approx$500–600\,d after the initial trigger, based on a rapid decline in the X-ray flux by several orders of magnitude \citep{Zauderer2013,Mangano2016,Eftekhari2018}. Some ulGRBs have been associated with supernovae, such as GRB 111209A \citep{Greiner2015} from which a super-luminous Type Ic supernova was detected, demonstrating the diversity in progenitors within the small known population of ulGRBs. 

Distinguishing between progenitors for ulGRBs remains a challenge.
Collapsar GRBs are typically characterised by shorter-lived central engines ($\lesssim$ few thousand seconds), highly variable prompt emission \citep[although some GRBs with supernova associations do not show highly variable prompt emission, e.g. GRB 060218][]{Campana2006}, and locations in star-forming regions offset from galaxy centres \citep{Fruchter2006}. Whereas, TDEs are often identified by their longer-lived accretion signatures, extended X-ray activity, and nuclear or pseudo-nuclear locations within their host galaxies \citep{Bloom2011,Burrows2011}. Additional constraints can be obtained from the temporal evolution of the afterglow or evidence for sustained energy injection \citep{Sari1998,Bromberg2012,Levan2011}. Convincing evidence to distinguish between the two, such as a supernova association \citep[e.g.][]{Greiner2015} or late-time jet shut-off \citep[e.g.][]{Zauderer2013,Eftekhari2018}, are often difficult to observe at high ($z\gtrsim1$) redshifts.

GRB 250702B is an unusual transient first detected in July 2025 and observed by numerous high-energy instruments on satellites, including the \textit{Fermi} Gamma-ray Burst Monitor \citep[GBM;][]{Fermi2009}, the \textit{Space Variables Objects Monitor} \citep[SVOM;][]{SVOM2011}, the \textit{Neil Gehrels Swift Observatory} \citep{Swift2004}, the \textit{Monitor of All-sky X-ray Image} \citep[MAXI;][]{Maxi2009}, Konus-Wind \citep{Konus1995}, and the \textit{Einstein-Probe} \citep[EP;][]{EP2022}. The prompt gamma-ray emission detected represents the longest GRB detected to date, causing three Fermi-GBM triggers (GRB 250702B/D/E) over $\approx3$\,hr on 2025-07-02 \citep{Neights2025GCN}.  Hard X-ray emission was detected by Konus-Wind coincident with the GRB emission, constraining a duration of $\gtrsim4.3$\,hr \citep{Frederiks2025GSN}. Curiously, EP detected X-ray emission one day prior to the Fermi GRB triggers on 2025-07-01 \citep{EP2025GCN,Zhang2026}, suggesting an extremely long-lived central engine. The prompt emission exhibited a hard spectrum with sub-second variability \citep{Neights2026,Oganesyan2025}. X-ray observations showed prolonged emission lasting weeks ($>65$\,d observer frame), with a steep decay ($F_X\propto t^{-1.7} - t^{-1.9}$) and strong short-timescale variability ($\Delta T/T<0.03$) superposed on the overall evolving light curve \citep{OConnor2025}.
Follow-up images with the Hubble Space Telescope revealed a galaxy with complex, asymmetric morphology, and an off-nuclear location of the transient \citep{Carney2025,Levan2025,Sears2026}. JWST spectroscopy measured a host redshift of $1.036\pm0.004$ \citep{Gompertz2026}, constraining the isotropic energy release in gamma-rays to at least $2.2\times10^{54}$\,erg and revealing a massive, extremely dusty host galaxy, unique among GRB hosts \citep{Gompertz2026}. Recent late-time JWST observations found tentative ($\sim3\sigma$) evidence for transient emission in the near-IR, indicative of a late-time light curve flattening \citep{Sears2026}. \citet{Sears2026} deduce that such a late-time flattening is consistent with either a TDE-like plateau or a supernova plus GRB afterglow model. 

The $>25,000$\,s gamma-ray emission observed from GRB 250702B is incompatible with standard (simple) GRB collapsar central engine models, which cannot sustain an accretion time greater than a few thousand seconds due to the physical limit of angular momentum available in a single star spinning at break-up velocity \citep{Woosley1993,MacFadyen1999,Fryer2025}. Therefore, more exotic progenitor scenarios have been invoked. The off-nuclear location of the transient within the host galaxy rules out classical nuclear SMBH TDEs. Current leading scenarios for the progenitor system include a He star merger \citep{Neights2026} or a TDE from either an intermediate mass black hole (IMBH) with a stellar \citep{Granot2026} or white dwarf \citep{Eyles2026,Sato2026,Yuan2026} companion, or from a stellar-mass black hole \citep{Beniamini2026}.
Initial multiwavelength modelling of the afterglow favours synchrotron emission from a forward (and possibly reverse) shock propagating into a wind-like ambient medium, consistent with an angular structured, relativistic outflow, although limited radio data have been used to date in these afterglow models \citep{Levan2025,OConnor2025,Granot2026}. These initial models require an extremely narrow ($\theta\sim1$\,deg) jet opening angle \citep{OConnor2025,Granot2026}, although these models provided a poor fit to the limited radio data points available at the time. 
   
Evidently, GRB 250702B is an extreme and unusual relativistic transient that does not fit cleanly into existing categories of high-energy astrophysical events. The transient exhibits both GRB-like prompt emission (including short-timescale variability) and longer-lived, TDE-like high-energy behaviour (e.g. X-ray flaring and long-duration gamma-ray emission). Key distinguishing power may come from the long-lived radio afterglow where constraints on the total outflow energetics, evolutionary timescales, and environment can be obtained.

In this work, we present detailed radio and millimeter monitoring observations of GRB 250702B constraining the evolution of the synchrotron emission produced by this event from 6--356\,d (observer frame) post-burst.  In Section \ref{sec:observations} we describe the new observations and data reduction presented in this work, in Section \ref{sec:spec_modelling} we model the radio to mm light curves and the radio to mm spectra as a synchrotron afterglow. In Section \ref{sec:equi} we carry out an equipartition analysis of the synchrotron emission to infer properties of the jet and its environment, and in Section \ref{sec:discussion} we discuss the nature of the GRB in the context of its radio emission and compare its properties to other GRBs and TDEs. Finally in Section \ref{sec:summary} we provide concluding remarks. All calculations in this work assume flat cosmology parameters with $H_0=67.4$\,km\,s$^{-1}$\,Mpc$^{-1}$ and $\Omega_0=0.315$ \citep{Planck2020} and all relative time measurements for the GRB assume $t_0=60858.58061$ MJD \citep[i.e. the time of the first reported GRB trigger for this event corresponding to the time of GRB 250702B, noting that there were three \textit{Fermi} triggers, denoted GRB 250702 B, D, and E discussed in][]{Neights2026}. 

\section{Observations and Data Reduction}\label{sec:observations}

\subsection{VLA observations}
We observed the coordinates of GRB 250702B with the Karl G. Jansky Very Large Array (VLA) between 2025 July 08 and 2026 June 07 ($\delta t = 6-356$\,d; programs: 24B-222, PI J. Miller-Jones; 25A-109, PI: Y. Yao). All data were calibrated using standard procedures in the Common Software Astronomy Application \citep[\texttt{CASA}, v5.6.3;][]{CASA2022}, including the VLA pipeline. At all frequencies, 3C286 was used for flux density and bandpass calibration, while J1832--1035 (4--33\,GHz) and J1822-0938 (1--4\,GHz) were used for phase calibration. Observations were taken with the array in C-configuration (2025-07 to 2025-09), B-configuration (2025-09 to 2026-01), and A-configuration (2026-02 to 2026-06). Additional manual RFI excision was performed in \texttt{CASA} when necessary, and images of the target field were created using the CASA task \texttt{tclean}. When sufficient signal-to-noise was achieved, frequency bands were split into 0.5--2\,GHz sub-bands (depending on the observing bandwidth) for imaging. The flux density of the target was measured using the \texttt{CASA} task \texttt{imfit} by fitting an elliptical Gaussian fixed to the size of the synthesised beam. An additional 5$\%$ uncertainty was added in quadrature to the statistical flux density uncertainty to account for systematic uncertainty due to the flux density bootstrapping. All flux densities are reported in the Appendix in Table \ref{tab:all_fluxes}. 

\subsection{ALMA observations}

We observed GRB 250702B with the Atacama Large Millimeter/submillimeter Array (ALMA) over 12 epochs between 2025 July 09 and 2025 October 14, under project code 2023.1.01731.T (PI: J. Miller-Jones). Five epochs were taken in Band 6 (central frequency 233\,GHz), ending in late August, and 7 epochs were taken in Band 3 (central frequency 97\,GHz). Table~\ref{tab:all_fluxes} gives the observing dates and times. Data were processed in \texttt{CASA} \citep{CASA2022} v6.6.1.17 using the ALMA pipeline (version 2024.1.0.8 for the first five epochs, and 2024.1.0.35 for the final two epochs). Data were taken with 8\,GHz of bandwidth at both frequency bands, using J1846$-$0651 as a phase calibrator. We used J1902$-$0458 as a check source at 97\,GHz, and J1834$-$0301 at 233\,GHz. In the earliest epochs (prior to 2025 July 28), when the source was sufficiently bright, the data were self-calibrated by the ALMA pipeline in each of the observing bands. Images were made with Briggs weighting, typically adopting a robust factor of 0.5, except in the final few epochs, when we adopted natural weighting to maximise the signal-to-noise ratio. Flux densities were measured by fitting an elliptical Gaussian in the image plane, whose size was fixed to that of the synthesized beam. An additional 5$\%$ uncertainty was added in quadrature to the statistical flux density uncertainty to account for systematic uncertainty due to the flux density bootstrapping. For the faintest epochs, we needed to use a forced fit, fixing the source position to the known value from the first (and brightest) epoch, which was (J2000) 18$^{\rm h}$58$^{\rm m}$45.5670$\pm$0.0007$^{\rm s}$, $-07^{\rm d}$52$^{\prime}$26.30$\pm$0.01$^{\prime\prime}$. Positional uncertainties were dominated by systematic uncertainties, determined as 10\% of the size of the synthesized beam from the first (brightest) epoch.

\subsection{uGMRT observations}
We obtained uGMRT observations of the field of GRB 250702B in two frequency bands - band 4 (central frequency 650\, MHz, bandwidth 400\,MHz) and band 5 (central frequency 1260\,MHz, bandwidth 400\,MHz),  between 2025 July 12 and 2025 September 25 (uGMRT ToO proposal 48\_059, PI: D. Eappachen). 
The raw data were downloaded in the \textsc{fits} format and converted to the \texttt{CASA} measurement set format. Then the data were calibrated and imaged using the automated continuum imaging pipeline \texttt{casa-capture} \citep{Kale2021}. Eight rounds of self-calibration were performed within each pipeline run. We detect a significant radio source at the position of the optical counterpart of GRB 250702B in all epochs. We compute the radio fluxes using the \texttt{CASA} task \texttt{imfit} in a small, circular region of radius approximately equal to the synthesized beam width, centered on R.A. = 18h58m45.5s, Dec. = $-$07d52m28.2s. 
An additional 30$\%$ systematic uncertainty was added in quadrature to the statistical flux density uncertainty to account for additional systematic uncertainty due to offsets in flux density measured between the VLA 1.25\,GHz and GMRT 1.26\,GHz measurements of multiple compact background sources in the field. 

\subsection{Published radio observations}
Numerous public research notes and Astronomer's Telegrams have been published on this source with a combination of archival and new radio observations. We supplement our VLA, GMRT, and ALMA datasets with radio observations from these research notes that are constraining. These include MeerKAT observations at 1.28 and 3\,GHz reported in \citet{Bright2025}, \citet{MeerKAT_ATel}, and \citet{Grollimund2025}. Additionally, the coordinates of GRB 250702B were serendipitously observed by MeerKAT $\approx2000$\,d (observer frame) prior to the transient occurring, placing a 3$\sigma$ upper bound of $\lesssim114\,\mu$Jy/bm at 1.28\,GHz on any host galaxy component to the radio emission \citep{MeerKAT_ATel}. 

\section{Results}\label{sec:spec_modelling}

\begin{figure}
    \centering
    \includegraphics[width=1.1\linewidth]{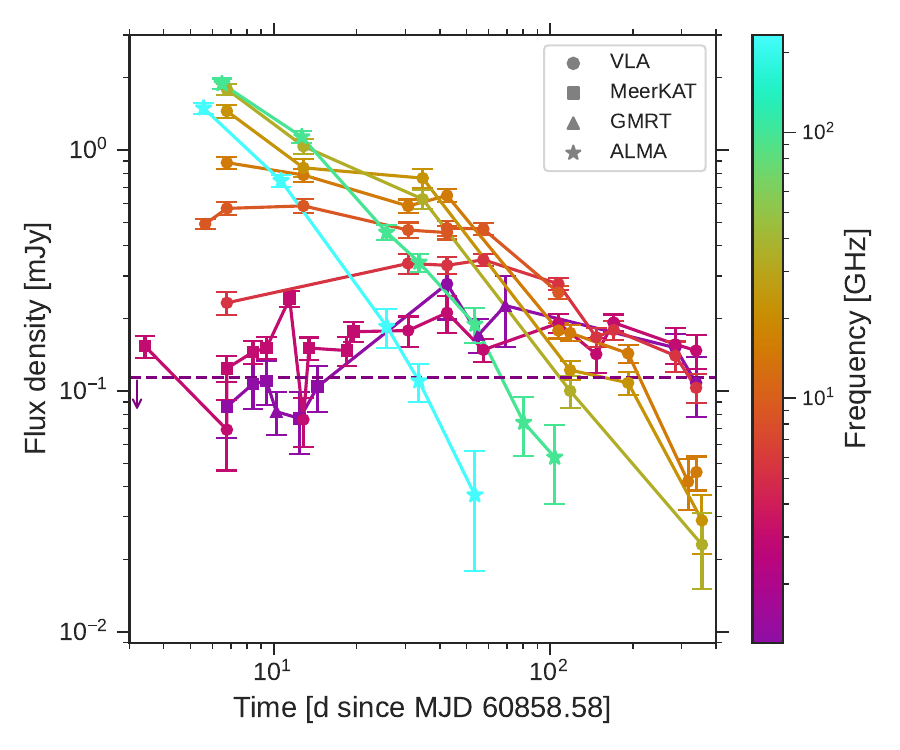}
    \caption{The observed radio light curves of GRB 250702B at 1.25, 3, 6, 10, 15, 22, 33, 97.5, and 233\,GHz. Different markers indicate different observatories: circles are VLA, squares are MeerKAT, triangles are GMRT, and stars are ALMA. The horizontal dashed purple line shows the 3$\sigma$ 1.28\,GHz upper limit on the host galaxy emission from a MeerKAT observation 2000\,d (observer frame) prior to the transient occurring. The higher frequencies ($>20$\,GHz) show a rapid decay, whereas the lower-frequency light curves break at later times.
    }    \label{fig:radio_lightcurves}
\end{figure}

The observed radio light curves for GRB 250702B are plotted in Figure \ref{fig:radio_lightcurves}. A rapid decay in flux density was observed from the initial observation at 6\,d post-burst at high ($>20$\,GHz) frequencies, while a significantly slower and flatter evolution was observed at lower frequencies.

\subsection{Light curve modelling}\label{sec:lightcurve_fits}

The temporal evolution of the light curve at different frequencies within the radio band can provide information about the mechanism powering the radio emission. In this section, we fit the observed radio light curves from 1.25--233\,GHz with a model-agnostic free broken power-law, in order to constrain the light curve properties across the different observed frequencies. We adopt a smoothly broken power-law of the form

\begin{equation}\label{eq:lc}
    F_{\nu} = F_{\nu,b} \left[\frac{1}{2}\left(\frac{t}{t_b}\right)^{-s\alpha_1} + \frac{1}{2}\left(\frac{t}{t_b}\right)^{-s\alpha_2} \right]^{- 1/s},
\end{equation}
where $F_{\nu}$ is the flux density observed at time $t$, $t_b$ is the break time, $\alpha_1$ the power-law index of the pre-break light curve, $\alpha_2$ the power-law index of the post-break light curve, and $s$ is a smoothing parameter which sets the sharpness of the break. In all fits we assume $s=2$ (upon testing different values this parameter had little impact on the fits obtained) and enforce $\alpha_1>\alpha_2$ to ensure the pre- and post-break powerlaw indices are easily distinguishable. 

We fit individual observed light curves at 1.25, 3, 5.5, 9, 15, 22, 34, 97, and 233\,GHz, combining flux densities if they are within 0.5\,GHz of the listed frequency between epochs. We use the Python implementation of MCMC, \texttt{emcee} \citep{emcee2013}, employing a Gaussian likelihood function. We allow the parameters to vary within the range $-1 < \log_{10}F_{\nu,b} \rm{[mJy]} < 10$, $0.1 < t_b \rm{[d]} < 400$, $-5 < \alpha_1 < 5$, and $-5 < \alpha_2 < 0$. We run each chain for 2000 steps with 200 walkers, discarding the first 1000 steps for burn-in. For each MCMC chain, we checked for convergence by measuring the $\hat{R}$ \citep[e.g.][]{Vehtari2021} using the \texttt{ArviZ} Python package, which in all cases was $<1.01$, and ensuring the chains were run for more than 50 times the \texttt{emcee} estimated autocorrelation time. 

\begin{figure*}
    \centering
    \includegraphics[width=\linewidth]{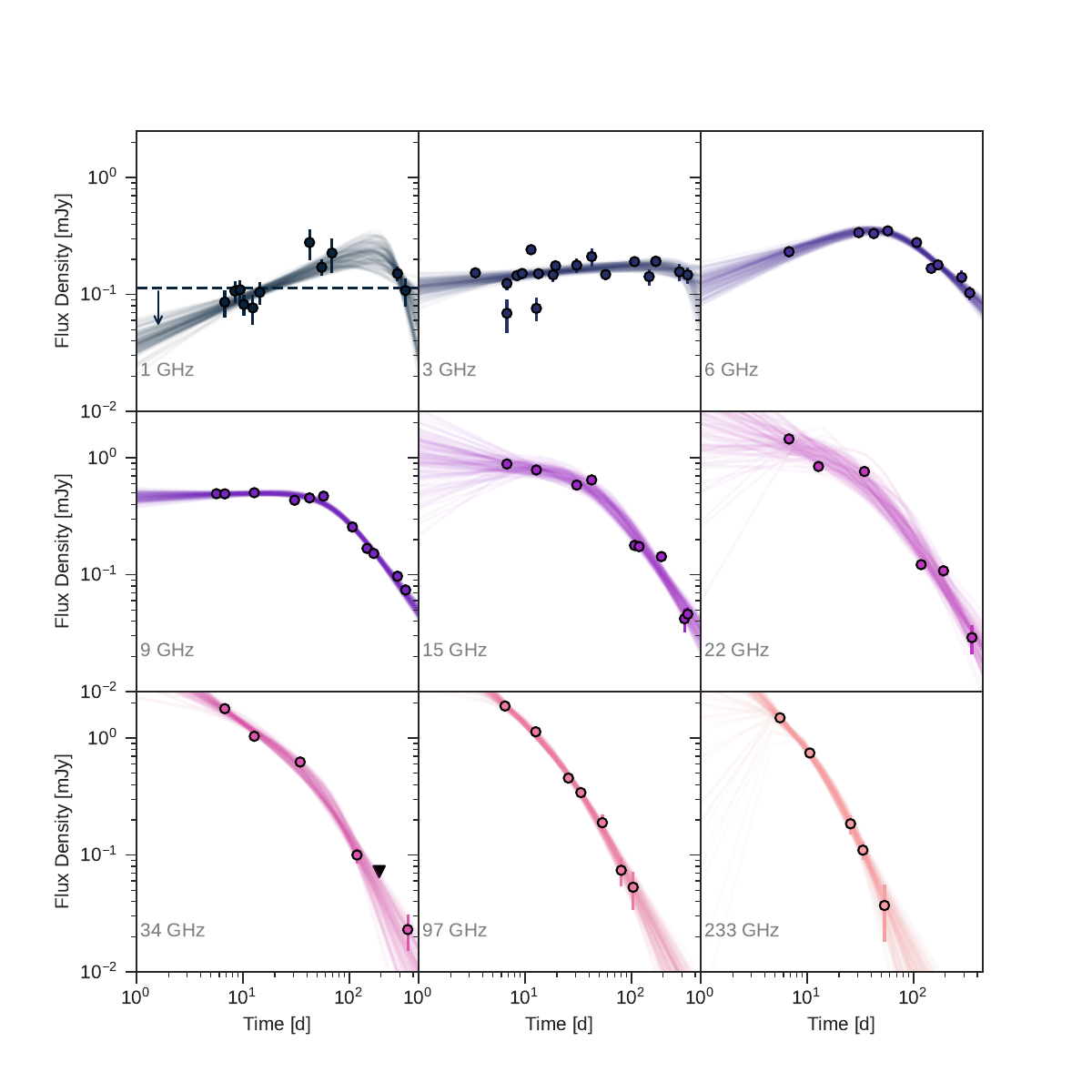}
    \caption{Broken power-law fits to the observed light curves from 1--233\,GHz. Inverted triangles indicate 3$\sigma$ upper limits. The horizontal dashed line in the top left panel shows the 3$\sigma$ upper limit on any host contribution to the flux density from archival MeerKAT observations. A clear frequency-dependence is observed in the power-law break times and decay indices.}
    \label{fig:radio_lightcurves_model}
\end{figure*}

\begin{figure*}
    \centering
    \includegraphics[width=\linewidth]{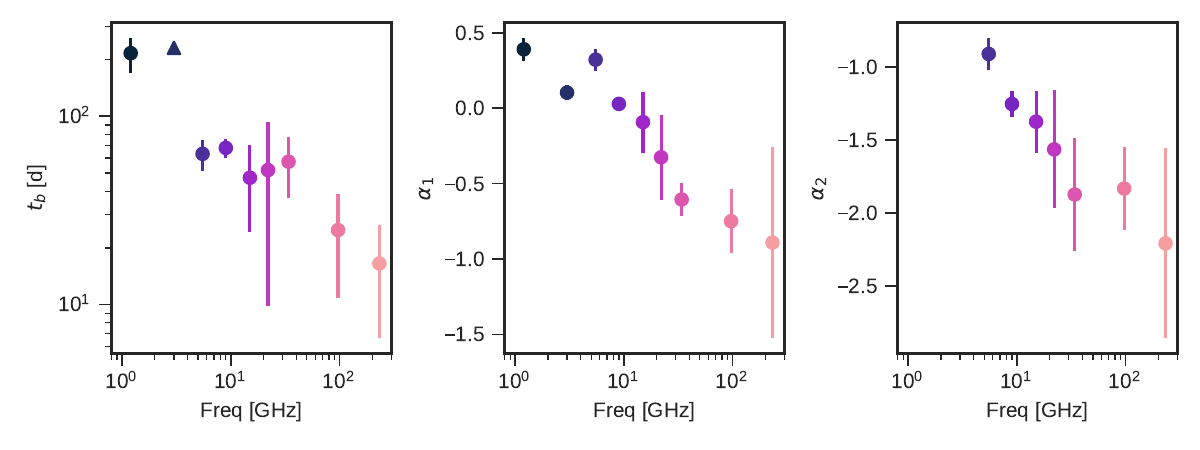}
    \caption{The power-law fit parameters for each frequency modelled in Figure \ref{fig:radio_lightcurves_model}. We see the break time of the radio light curves is strongly frequency-dependent, as are the pre- and post-break power-law indices.}
    \label{fig:radio_lightcurves_fitparams}
\end{figure*}

The resulting light curve fits are plotted in Figure \ref{fig:radio_lightcurves_model} and the best-fit parameters for each frequency are plotted in Figure \ref{fig:radio_lightcurves_fitparams} and listed in Table \ref{tab:lc_fits}. A break time is constrained at all frequencies except 3\,GHz. The light curve was decaying at all frequencies $>15$\,GHz from the beginning of our radio campaign at 6\,d post-burst (observer frame), with the light curves at those frequencies observed to break to a steeper decay-index at different times. At frequencies $\lesssim15$\,GHz, the light curve is initially flat or rising, before breaking to a steeper decay. The break times in general increase with decreasing frequency, as do the pre- and post-break spectral indices. Significant variability around the best-fit model is observed at 1.25 and 3\,GHz, which we discuss in the following sections. 

\begin{table}[]
    \centering
    \begin{tabular}{ccccc}
   Freq.(GHz) & $F_{b}$ (mJy) & $t_b$ (d) & $\alpha_1$ & $\alpha_2$ \\
   \hline
   \hline
1 & $0.21\pm0.04$ & $220\pm42$ & $0.38\pm0.08$ & -- \\
3 & $0.15\pm0.02$ & $>231$ & $0.10\pm0.05$ & -- \\
6 & $0.33\pm0.02$ & $64\pm12$ & $0.32\pm0.07$ & $-0.9\pm0.1$ \\
9 & $0.37\pm0.02$ & $68\pm8$ & $0.03\pm0.03$ & $-1.3\pm0.1$ \\
15 & $0.50\pm0.17$ & $49\pm22$ & $-0.11\pm0.18$ & $-1.4\pm0.2$ \\
22 & $0.61\pm0.57$ & $37\pm46$ & $-0.29\pm0.44$ & $-1.3\pm0.5$ \\
34 & $0.24\pm0.06$ & $57\pm20$ & $-0.61\pm0.11$ & $-1.9\pm0.4$ \\
97 & $0.50\pm0.31$ & $25\pm14$ & $-0.76\pm0.19$ & $-1.8\pm0.3$ \\
233 & $0.36\pm0.41$ & $18\pm9$ & $-0.94\pm0.34$ & $-2.3\pm0.6$ \\
    \hline
    \end{tabular}
    \caption{Light curve fit parameters for individual frequencies fit with Equation \ref{eq:lc}}
    \label{tab:lc_fits}
\end{table}

The trend across the observed frequencies for the light curve break time to increase with decreasing frequency provides crucial insight into the mechanism powering the observed radio emission. It is immediately clear in our light curve fits that the timing of the observed breaks is not achromatic, implying the shift in evolution is likely due to critical frequencies sweeping through the band rather than a jet-break or other dynamical effect. 

Using standard closure relations, the light curve evolution can be related to the synchrotron electron index, $p$ (where the electrons are accelerated into a powerlaw distribution $N(\gamma)\propto \gamma^{-p}$ where $\gamma$ is the electron Lorentz factor), and the density stratification of the environment, $k$ (where the ambient density, $n_e$ drops off with radius $R$ as $n_e\propto R^{-k}$), depending on the regime and physics of the shock \citep[e.g.][]{Gao2013}. The diversity in the observed light curve fit indices (Table~\ref{tab:lc_fits}) is likely indicative that different radio bands sample different synchrotron spectral regimes, with critical frequencies potentially evolving through the bands. This conclusion is further supported by the strong frequency dependence of the lightcurve break times. At low frequencies (1--9\,GHz), the pre-break slopes are positive or near zero ($\alpha_1 \simeq 0.0$ to $0.4$), indicating that these bands lie below the synchrotron peak at early times. The 15--22\,GHz bands are nearly flat or only mildly declining before the break, consistent with sampling the transition as $\nu_m$ and/or $\nu_a$ moving through the band. By contrast, the higher-frequency light curves (34--233\,GHz) are already decaying prior to the break ($\alpha_1 \simeq -0.6$ to $-0.9$), implying that these frequencies lie above $\nu_m$ at early times.

After the break, the best-constrained bands (5.5--22\,GHz) show decay slopes of $\alpha_2 \simeq -0.9$ to $-1.3$, broadly consistent with optically thin forward-shock emission in the regime $\nu_m < \nu < \nu_c$ \citep{Gao2013}. A decay index of $-1$ to $-2$ implies $p\approx2-3$, consistent with the spectral fits that find $p=2.5$ (Appendix \ref{sec:p_fits}). However, the steepest post-break decays, seen in the 34, 97, and 233\,GHz light curves, are steeper than expected for a simple forward-shock model.

\subsubsection{Low-frequency excess emission: a host component?}
The low-frequency light curves at 1 and 3\,GHz show very flat evolution, particularly at early times post-burst. The 1.25\,GHz data has a weighted mean flux density of 0.1\,mJy. A flux density of $\approx0.1$\,mJy at 1.25\,GHz for the host galaxy is not ruled out by the archival MeerKAT radio observations, which provided a $3\sigma$ upper limit of $<0.114$\,mJy at 1.28\,GHz \citep{MeerKAT_ATel}.
However, a $\chi$-squared test that the flux density is constant across the 1.25\,GHz measurements yields a reduced $\chi$-squared value of 2.23, with p-value of 0.014, indicating a 2.5$\sigma$ significance that the light curve is not constant. At 3\,GHz, the weighted mean is 0.153\,mJy. If all of the observed emission was from the host galaxy at 1 and 3\,GHz this would imply an extremely unusual host galaxy spectrum with a positive spectral index, unlike typical AGN or star-forming galaxies \citep{Murphy2011,Gordon2021}. A $\chi$-squared test that the flux density is constant across the 3\,GHz measurements yields a reduced $\chi$-squared value of 4.68 and a p-value of $2\times10^{-10}$, rejecting the null hypothesis at approximately 6.3\,$\sigma$ significance. This result indicates there is likely a transient component to the measured flux density at both frequencies, but particularly at 3\,GHz.  

We therefore caution that some of the low-frequency $\lesssim1.5$\,GHz radio emission from this source may be due to host galaxy emission as we cannot rule this out and the light curve behaviour is unusually flat for pure synchrotron emission. We therefore do not include the $<1.5$\,GHz flux density measurements in our spectral fitting; these data are not crucial to the conclusions of this work.

\subsubsection{Interstellar Scintillation}\label{sec:disc_flux_excess}

There is a large amount of somewhat stochastically appearing variability in the observed 0.65, 1.25, and 3\,GHz light curves at early times (Figure \ref{fig:radio_lightcurves}). This variability could be due to interstellar scintillation (ISS). ISS can cause significant modulation of the radio flux density observed from a compact extragalactic radio source, depending on the electron density along the line of sight and intrinsic source size. Previous studies of GRBs have found evidence for ISS induced variability, which can sometimes be used to place upper limits on the radio-emitting region size \citep[e.g.][]{Goodman1997,Frail1997,Chandra2008,vanderhorst2014,Alexander2017,Anderson2023}. 

We implement the ISS formalism derived by \citet{Walker1998} as appropriate for compact extra-galactic radio sources to infer the scintillation properties at different observing frequencies and source sizes. These calculations assume the bulk of the electrons causing the ISS are due to electrons within our Galaxy, although we note there is an unknown and unquantifiable distribution of electrons in the host galaxy that the radio emission is also likely passing through before reaching our detectors. This contribution may be particularly important for the host galaxy of GRB 250702B, which is atypically dusty \citep{Gompertz2026}, however, the contribution from electrons in the host galaxy is unquantifiable in our analysis. We further note that for extragalactic sources, the Milky Way electron density models (which are primarily calibrated using pulsars which lie in the plane of the Galaxy) can be inaccurate, and in some cases variability in excess of the predicted ISS modulation has been observed \citep[e.g. GRB 161219B;][]{Alexander2019grb}. Therefore, the constraints on the source size obtained via this scintillation analysis are potentially subject to large systematic error and should be considered order of magnitude estimates.

The observational manifestation of ISS for a point source is modulation of the radio flux density characterised by a modulation index (i.e. the rms fractional flux variation) and corresponding timescale of modulation.
Following \citet{Walker1998}, in the strong, refractive scintillation regime the modulation index is given by 
\begin{equation}
    m_p = \left(\frac{{\nu_{\rm{obs}}}}{\nu_0}\right)^{17/30}
\end{equation}
where $nu_0$ is the critical transition frequency and the timescale of scintillation is given by
\begin{equation}
    t_{\rm{scint}} = 2\left(\frac{\nu_{\rm{0}}}{\nu_{\rm{obs}}}\right)^{11/5} \rm{[hr]}
\end{equation}
if the source angular size $\theta_s$ is less than the scattering source size,

\begin{equation}
    \theta_{\rm{scatt}} = \theta_F \left(\frac{\nu_0}{\nu_{\rm{obs}}}\right)^{11/5} [\mu\rm{as]}
\end{equation}
where $\theta_F$ is the angular size of the first Fresnel zone.

In the scenario in which the source size $\theta_s > \theta_{\rm{scatt}}$, the modulation index is reduced by a factor $(\theta_{\rm{scatt}}/\theta_s)^{7/6}$ and the timescale for variability increases as $\theta_s/\theta_{\rm{scatt}}$. We only consider refractive scintillation in this work and not diffractive scintillation due to the integration times of our observations; diffractive scintillation applies on timescales significantly shorter than our observation integration time and so should be smeared out.

At the Galactic coordinates of GRB 250702B, the NE2025 electron density model \citep{Ocker2026} predicts a transition frequency between strong and weak scintillation regimes of 6.7\,GHz and angular size of the first Fresnel zone of $\theta_F=0.5$\,$\mu$as. 

\begin{figure*}
    \centering
    \includegraphics[width=\linewidth]{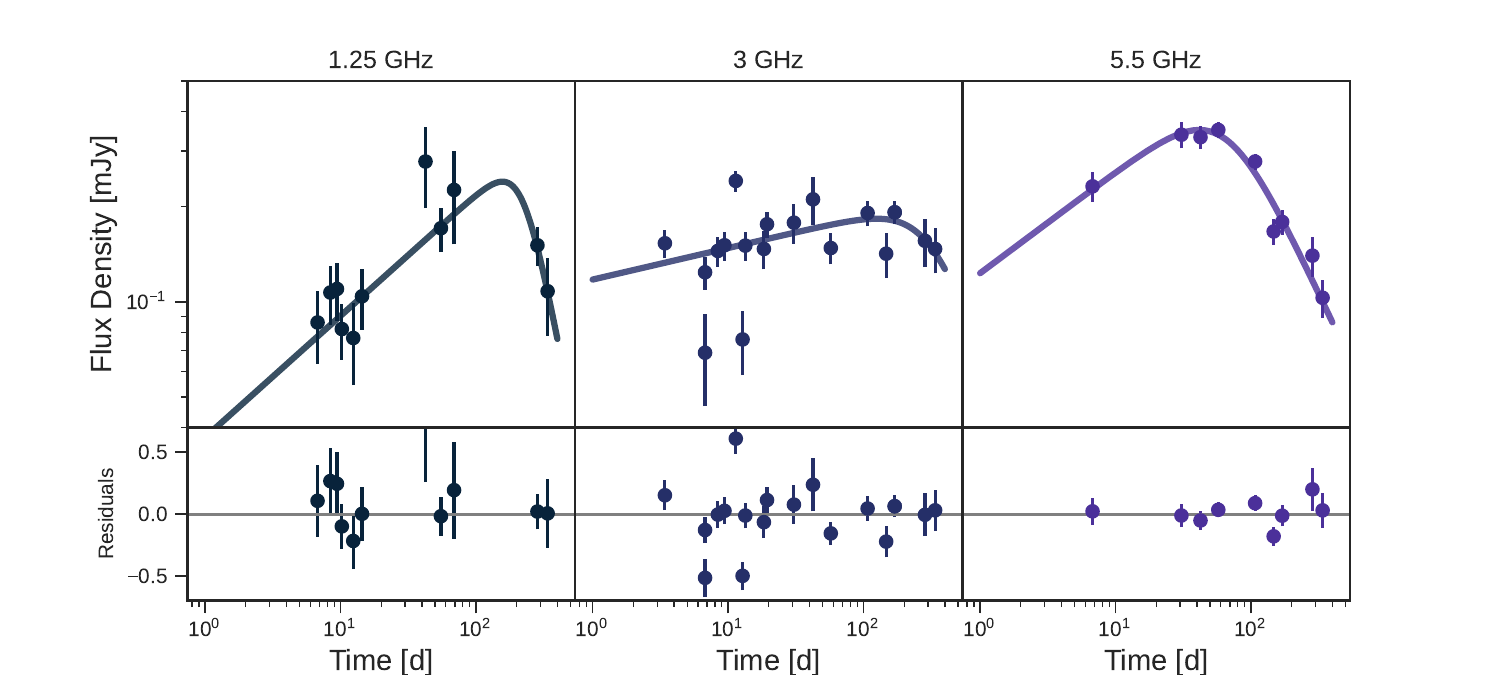}
    \caption{The 1.25, 3, and 5.5\,GHz light curves with the fractional model residuals plotted underneath (Residual = $(F_{\rm{obs}} - F_{\rm{model}}) / F_{\rm{model}}$). Significant variability is observed at 3\,GHz, particularly at early times. The largest residuals are $0.8\pm0.5$ at 1.25\,GHz, $0.6\pm0.1$ at 3\,GHz, and $0.2\pm0.2$ at 5.5\,GHz. We interpret this low frequency variability as a signature of interstellar scintillation.}
    \label{fig:scintillation}
\end{figure*}

To quantify any excess variability around the broken power-law model that could potentially be due to ISS, we first calculate the fractional residual of the flux densities using

\begin{equation}
    r_i = \frac{F_{\rm{obs},i} - F_{\rm{model},i}}{F_{\rm{model},i}}
\end{equation}
where $F_{\rm{model}}$ is the flux density from the broken power-law fit to the data in Section \ref{sec:lightcurve_fits}. These residuals are plotted in Figure \ref{fig:scintillation}. Note that we do not include the 0.65\,GHz data in the variability analysis as there are only 3 measurements available at that frequency.

Whilst the usual approach to measuring the modulation index in scintillation analyses is to take the mean of the observations and calculate the root-mean-square fractional flux variation \citep{Walker1998}, since there is intrinsic variability of the flux density observed for this source (well above the expectations of variability due to ISS), we instead measure the modulation index of the model residuals as a proxy for measuring the excess variability above the model. This approach assumes the radio lightcurves follow the smoothly-broken power-law model fit in Section \ref{sec:lightcurve_fits}
Therefore, to estimate the modulation index of the fractional residuals we take the residual standard error, i.e.

\begin{equation}
    m_{\rm{p,obs}} = \sqrt{\frac{1}{N}\sum{\rm{r_i}^2}}
\end{equation}

\noindent where $N$ is the number of data points.

In order to account for uncertainty in both the model and observed data used to measure the residuals, we randomly sample $F_{\rm{obs},i}$ within the error bars of each data point and randomly sample a single $F_{\rm{model}}$ from the highest 50$\%$ likelihood walkers of the MCMC fit, to generate a distribution of $m_{\rm{p, obs}}$ for 100,000 samples. We report the median and standard deviation of this residual distribution in Table \ref{tab:scintillation}. This approach ensures that the estimated modulation fraction accounts for error in both the data and the model, allowing an estimate of $m_{\rm{p, obs}}$ and an error to be obtained. 

At an observing frequency of 1.25\,GHz, the source will be in the strong scintillation regime until it reaches a source size of $\theta_{\rm{scatt}}=20$\,$\mu$as. In the strong scintillation regime, we expect a scintillation modulation fraction of 0.39 on a timescale of 80\,hrs at 1.25\,GHz. At an observing frequency of 3\,GHz, the source will be in the strong scintillation regime until it reaches a source size of  $\theta_{\rm{scatt}}=3$\,$\mu$as. In the strong scintillation regime, we expect a scintillation modulation fraction of 0.63 on a timescale of 11.7\,hrs at 3\,GHz. In contrast, at frequencies above the critical transition frequency of 6.7\,GHz, we expect the source to be in the weak scintillation regime and very little modulation to be observed in the flux density. This lack of variability at the higher frequencies is consistent with the observed multi-frequency light curve (Figure \ref{fig:radio_lightcurves} and \ref{fig:scintillation}). We summarise the predicted and observed scintillation properties at the three frequencies considered in Table \ref{tab:scintillation}.

In Figure \ref{fig:scintillation}, we plot the model residuals for the 1.25, 3, and 5.5\,GHz light curves, quantifying the larger variability around the model observed at 1.25 and 3\,GHz than at higher frequencies. The level of frequency-dependent variability is entirely in keeping with the expectations of ISS, and therefore requires that the source size is small enough so as to not be completely resolved to ISS at the lower observing frequencies. The 3\,GHz modulation fraction is lower than the theoretical modulation fraction if the source was completely unresolved to scintillation. Instead, we can adjust the modulation fraction and obtain a constraint on the average source size at 3\,GHz based on the level of scintillation observed. We find an observed modulation fraction of 0.3 implies $\theta_s\approx5.67\,\mu$as. At 1.25\,GHz, the observed modulation fraction is consistent within error of the theoretical modulation fraction, suggesting the source is unresolved to scintillation and placing an upper bound on the source size of $\theta_s\lesssim20\,\mu$as. Whereas, at 5.5\,GHz, very little modulation is observed, and significantly less than the theoretical modulation fraction of 0.89 (since the observing frequency is so close to the transition frequency). This therefore implies that the source is mostly resolved to scintillation at 5.5\,GHz, suggesting a lower limit on the source size of $\theta_s\gtrsim0.77\,\mu$as. 

Further evidence for ISS is seen in VLA and MeerKAT observations taken on the same day (2025-07-09) 2.5 hours apart, in which we observed the flux density of the source to change from $60\pm11$\,$\mu$Jy to $124\pm15$\,$\mu$Jy at a central observing frequency of 3\,GHz. Within the $1\sigma$ error bars, this requires at minimum a 17$\%$ change in flux density over this 2.5\,hr period, but the errors allow a modulation fraction up to $>50\%$. This variability is significantly larger than any expected flux density calibration offset between the two instruments (usually assumed to be $\approx5\%$), and given the source is point-like and unresolved the differing UV-coverage of the two instruments is unlikely to produce such a large offset in flux density. 

ISS provides a constraint on the maximum size of the jet front image on the sky, which we assume is a proxy for the blastwave image in the plane of the sky \citep{Granot2005}. If the source is unresolved to ISS, the blastwave image in the plane of the sky cannot be significantly larger than the physical size of the scattering disk, i.e. $R_{\perp}\approx\theta_{\rm{scatt}}D_A$, where $D_A$ is the angular size distance to the source (at $z=1.036$, $D_A=5.288\times10^{27}$\,cm). In Table \ref{tab:scintillation} we also provide the constraints that the observed level of scintillation implies for $R_{\perp}$. We emphasise that these source size estimates should be considered an order of magnitude estimate due to the numerous assumptions made in predicting and observing the scintillation properties. 

\begin{table}
    \centering
    \begin{tabular}{l|l|lll|ll} 
    $\nu_{\rm{obs}}$ & $m_{\rm{p,obs}}$ & $\theta_{\rm{scatt}}$ & $m_p$ & $t_{\rm{scint}}$ & $\theta_{source}$ &$R_{\perp}$ \\
    (GHz) & &($\mu$as) & & (hr) &($\mu$as) & (cm) \\
    \hline
    \hline
      1.25 & $0.3\pm0.1$ & 20 & 0.39 & 80 & $\lesssim20$ &$\lesssim5\times10^{17}$ \\
       3  & $0.26\pm0.02$ & 3 & 0.63 & 11.7 & $\approx5.67$& $\approx1.5\times10^{17}$\\
       5.5 & $0.12\pm0.03$ & 0.77 & 0.89 & 3 &$\gtrsim0.77$ &$\gtrsim2\times10^{16}$ \\
    \hline
    \end{tabular}
    \caption{The observed and predicted ISS properties at the given observing frequencies, and the resulting constraints on the blast wave image radius.}
    \label{tab:scintillation}
\end{table}

\subsection{Spectral modelling}\label{sec:spectral_fitting}

GRB 250702B shows an evolving, peaked spectrum with time. In this section we fit the observed spectra at each epoch with a physical synchrotron emission model to constrain the synchrotron properties of the spectrum. We use the synchrotron spectral model for GRB afterglows \citep{Granot2002} in which the synchrotron-emitting electrons are assumed to be accelerated into a power-law distribution, $N(\gamma_e)\propto \gamma_e^{-p}$. We assume slow cooling, motivated by the observed spectral shape, i.e. spectral fits assuming fast-cooling would be inconsistent with the observed spectral shape and slopes. The resulting spectrum has characteristic break frequencies corresponding to the minimum frequency ($\nu_m$), the self-absorption frequency ($\nu_a$), and the cooling frequency ($\nu_c$). The relative ordering of these break frequencies depends on the physics of the shock itself. For GRBs, usually at early times $\nu_a < \nu_m$ and the synchrotron spectrum is described by

\begin{equation}
    \label{eq:Fv1}
    \begin{aligned}
        F_{\nu, \mathrm{1}} = F_{\nu,\mathrm{p}} \left[
        \left(\frac{\nu}{\nu_{\rm a}}\right)^{-s_1\beta_1} +  \left(
        \frac{\nu}{\nu_{\rm a}}\right)^{-s_1\beta_2
        }\right]^{-1/s_1} \times \\
        \left[1 + \left(\frac{\nu}{\nu_{\rm m}}\right)^{s_2(\beta_2 - \beta_3)}\right]^{-1/s_2}
        \end{aligned}
    \end{equation}
    where $\nu$ is the frequency, $F_{\nu,\mathrm{p}}$ is the normalisation, $s_1 = 1.06$, $\beta_1 = 2$, $\beta_2 = \frac{1}{3}$, $s_2=1.76-0.38p$, and $\beta_3=\frac{1-p}{2}$. The values for $s_1$ and $s_2$ are appropriate for a wind $k=2$ environment.

In the adiabatic expansion phase, $\nu_m$ evolves more quickly with time than $\nu_a$, and the spectrum will transition to the regime in which $\nu_m < \nu_a$ for which the synchrotron spectrum is given by

\begin{equation}
    \label{eq:Fv2}
    \begin{aligned}
        F_{\nu, \mathrm{2}} = F_{\nu,\mathrm{p}} \left[\left(\frac{\nu}{\nu_{\rm m}}\right)^2 \exp(-s_3\left(\frac{\nu}{\nu_{\rm m}}\right)^{2/3}) + \left(\frac{\nu}{\nu_{\rm m}}\right)^{5/2}\right] \times \\
        \left[1 + \left(\frac{\nu}{\nu_{\rm a}}\right)^{s_4(\beta_4 - \beta_5)}\right]^{-1/s_4}
        \end{aligned}
    \end{equation}
    where $\nu$ is the frequency, $F_{\nu,\mathrm{p}}$ is the normalisation, $s_3 = 3.63p-1.60$, $s_4 = 1.25-0.18p$, $\beta_4 = \frac{5}{2}$, and $\beta_5 = \frac{1-p}{2}$.

\begin{figure*}
    \centering
    \includegraphics[width=\textwidth]{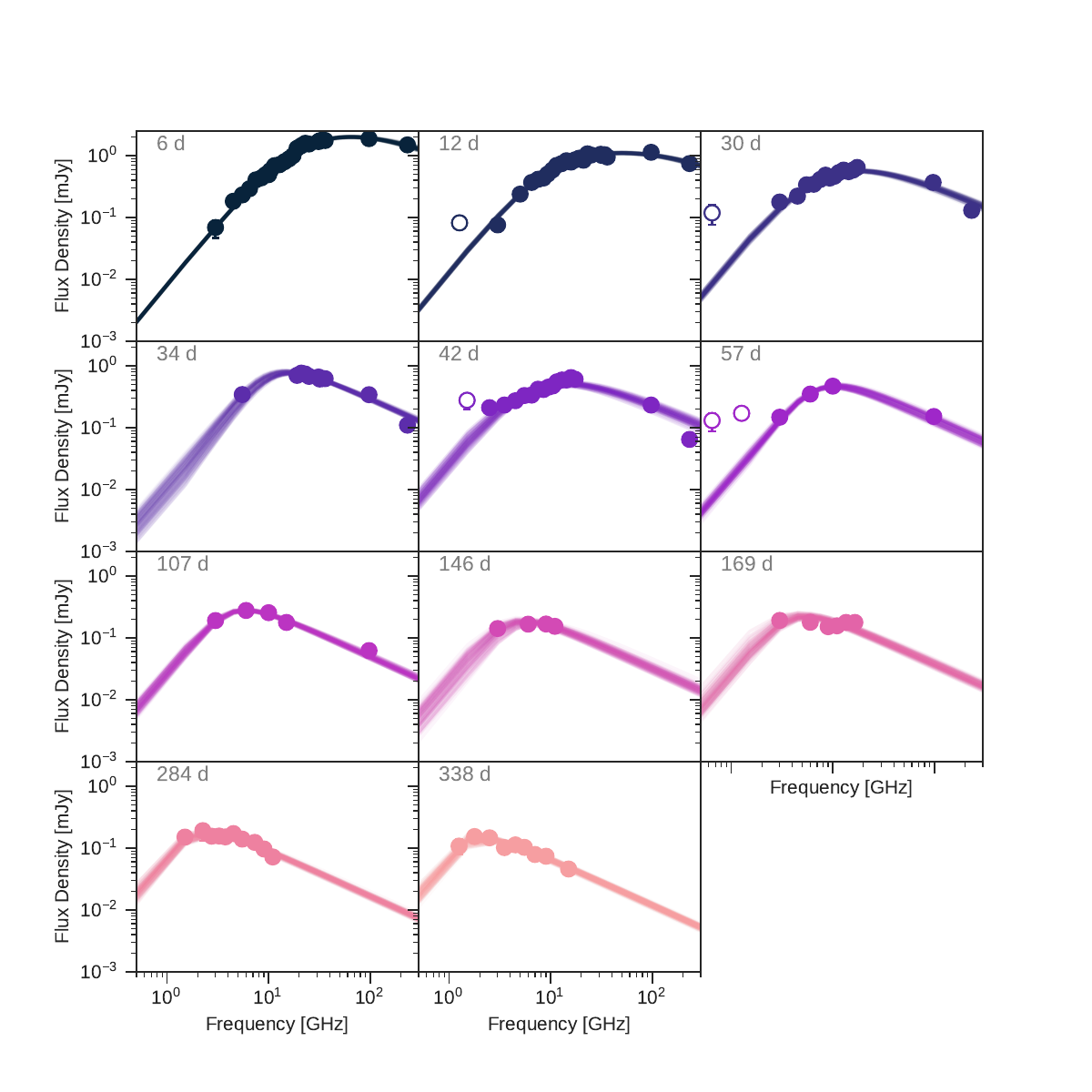}
    \caption{Synchrotron spectral fits to each epoch of broadband spectral observations of GRB 250702B from 0.65--233\,GHz. Open circles indicate measurements excluded from the spectral fitting, likely in excess due to ISS. Note that some of the 5.5, 97.5 and 233\,GHz flux densities have been extrapolated to obtain estimates of the flux density contemporaneous with the other frequencies, as described in Appendix \ref{sec:flux_extrapolation}.}
    \label{fig:spec_fits}
\end{figure*}

The exact ordering of the critical frequencies in our observations of GRB 250702B is uncertain.
In order to allow the observed spectrum to freely transition between these regimes in our fitting, we follow the methodology outlined in \citet{Eftekhari2018} and implement a weighting scheme to smoothly transition between the synchrotron regimes. Whilst this parametrisation is not physical, it is a practical solution to track the evolution of the critical frequencies as they evolve at different rates through the observing bands. We define a combined power-law flux, $F_{\rm{comb}}$, given by

\begin{equation}\label{eq:comb_Fv}
    F_{\rm{comb}} = \frac{w_1F_{\nu, \mathrm{1}} + w_2F_{\nu, \mathrm{2}} }{w_1 + w_2}
\end{equation}
where $w_1=(\frac{\nu_m}{\nu_a})^2$ and $w_2=(\frac{\nu_a}{\nu_m})^2$. 
We note that alternative approaches to model the free transition between thees two synchrotron regimes are explored in \citet{Leung2026b,Giarratana2026}.

To obtain constraints on the evolution of $F_{\nu,p}$, $\nu_m$, and $\nu_a$ over time, we fit 11 epochs of radio spectra with Equation \ref{eq:comb_Fv} using an MCMC approach. We allow $F_{\nu,p}$, $\nu_m$, and $\nu_a$ to freely vary in the range $-3<\log_{10}F_{\nu,p} [\rm{mJy}] < 2$, $-2 < \log_{10}\nu_m [\rm{GHz}] < 2.5$, and $-2 < \log_{10}\nu_a [\rm{GHz}] < 2.5$.  We fix $p=2.5$ in all fits (see Appendix \ref{sec:p_fits}). We implement a Gaussian log likelihood and use 400 walkers and 5000 steps, discarding the first 2000 steps for burning for each epoch. We additionally implement a likelihood selection on the remaining walkers to exclude walkers in low-probability regions of parameter space. For each MCMC chain, we checked for convergence by measuring the $\hat{R}$ \citep[e.g.][]{Vehtari2021}, which in all cases was $<1.001$, and ensuring the chains were run for more than 50 times the \texttt{emcee} estimated autocorrelation time. 

The resulting synchrotron spectral fits for each epoch are plotted in Figure \ref{fig:spec_fits}. We find in general the spectral model provides a good fit to the data, except the $<1.5$\,GHz measurements, for which a flux excess is observed. This excess could be due to ISS or weak host galaxy emission, discussed in Section \ref{sec:disc_flux_excess}. We therefore exclude flux density measurements $<1.5$\,GHz for the spectral fitting.

\begin{table}[]
    \centering
    \begin{tabular}{ccccc}
MJD & $t$ (d) & $F_p$ (mJy) & $\nu_a$ (GHz) & $\nu_m$ (GHz)   \\ 
\hline
\hline
60865 & $6$ & $3.37\pm0.54$ & $19.85\pm1.58$ & $72.55\pm14.13$ \\
60871 & $12$ & $1.21\pm0.09$ & $9.42\pm0.56$ & $85.43\pm10.20$ \\
60889 & $30$ & $1.59\pm0.25$ & $7.64\pm0.66$ & $12.92\pm1.91$ \\
60893 & $34$ & $1.31\pm0.63$ & $9.74\pm0.89$ & $8.20\pm1.49$ \\
60901 & $42$ & $1.52\pm0.26$ & $6.05\pm0.85$ & $9.33\pm1.77$ \\
60916 & $57$ & $1.02\pm0.22$ & $6.46\pm0.40$ & $5.95\pm0.68$ \\
60966 & $107$ & $0.44\pm0.22$ & $3.64\pm0.18$ & $3.00\pm0.56$ \\ 
61005 & $146$ & $0.32\pm0.14$ & $3.67\pm0.57$ & $3.03\pm0.74$ \\
61028 & $169$ & $0.44\pm0.15$ & $3.24\pm0.38$ & $<2.80$ \\
61143 & $284$ & $0.30\pm0.14$ & $1.55\pm0.16$ & $<1.27$ \\
61197 & $338$ & $0.23\pm0.11$ & $1.30\pm0.18$ & $<1.04$ \\
\hline
    \end{tabular}
    \caption{Synchrotron spectral fitting results for each epoch of radio spectral observations presented in this work.}
    \label{tab:spec_fits}
\end{table}

\begin{figure*}
        \centering
    \includegraphics[width=0.49\textwidth]{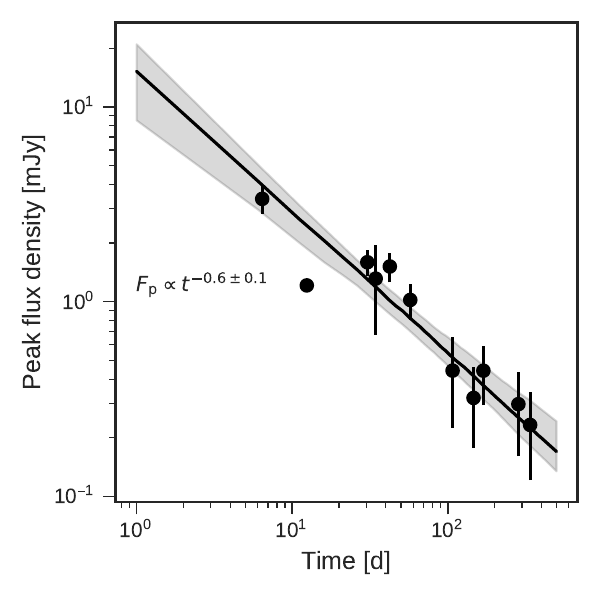}
        \includegraphics[width=0.49\textwidth]{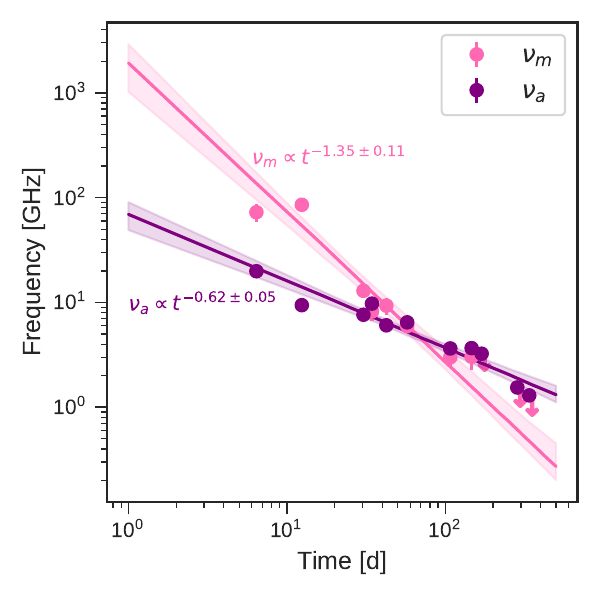}
    \caption{The evolution of the peak synchrotron flux density (left) and the synchrotron critical frequencies $\nu_a$ and $\nu_m$ (right). The shaded regions show $1\sigma$ confidence intervals from an MCMC power-law fit.}
    \label{fig:peaks_ev}
\end{figure*}

The inferred evolution of $F_{\nu,p}$, $\nu_m$, and $\nu_a$ over time is plotted in Figure \ref{fig:peaks_ev}. We fit $F_{\nu,p}$, $\nu_m$, and $\nu_a$ with a simple powerlaw of the form $At^b$ again using an MCMC approach and freely varying $A$ in the range $-3 < \log_{10}A [\rm{mJy\,or\, GHz}] < 4$ and $-10 < b < 0$. The fits are plotted in Figure \ref{fig:peaks_ev}. We find $F_{\nu,p}\propto t^{-0.6\pm0.1}$, $\nu_m\propto t^{-1.35\pm0.11}$, and $\nu_a\propto t^{-0.62\pm0.05}$.

The temporal evolution of the peak flux density, self-absorption frequency, and minimum frequency strongly disfavour a standard forward shock in a constant-density ($k=0$) medium, for which we would expect constant $F_{\nu,p}$ and $\nu_a$ with time in the adiabatic self-similar solution \citep{Gao2013}. By contrast, a stratified (e.g. $k=2$) ambient medium provides a much better description of the data. We find that the observed decay indices of all three spectral parameters ($F_p$, $\nu_a$, and $\nu_m$) are in good agreement with the theoretical expectation of a standard adiabatic forward-shock model \citep{Gao2013}. The observed peak flux density decay is slightly steeper than the theoretical $k=2$ forward shock solution ($F_{p,\rm{obs}}\propto t^{-0.6\pm0.1}$ compared to $F_{p,\rm{FS}}\propto t^{-0.5}$), the observed evolution of $\nu_a$ is consistent with the theoretical expectation ($\nu_{a,\rm{obs}}\propto t^{-0.62\pm0.05}$ compared to $\nu_{a,\rm{FS}}\propto t^{-0.6}$), and the observed evolution of $\nu_m$ is slightly flatter than expected ($\nu_{m,\rm{obs}}\propto t^{-1.35\pm0.11}$ compared to $\nu_{m,\rm{FS}}\propto t^{-1.5}$) \citep{Gao2013}. The observed peak flux density decay is much shallower than expected for a standard top-hat reverse shock propagating into a wind environment ($F_{p,\rm{obs}}\propto t^{-0.6}$ compared to $F_{p,\rm{RS}}\propto t^{-1.1}$) and the evolution of $\nu_m$ is again too shallow \citep{Gao2013}. However, some individual steep post-break light-curve slopes are fairly consistent with the (steeper) expectations of a reverse-shock decay (Table \ref{tab:lc_fits}). 

Typically, long-lived and slowly evolving radio afterglows are expected for forward-shock emission \citep[e.g.][]{Sari1998}, whilst reverse shock emission is expected to peak and fade quickly, i.e. within $\approx$hours at optical frequencies and $\approx$days at radio frequencies \citep{Sari1999b}. However, strong and relatively long-lived ($\sim10$s of days) reverse shock emission at radio frequencies has been detected in a growing number of GRBs \citep[e.g.][]{Laskar2013,Alexander2017,Zhang2024} and a combination of forward and reverse shock emission is often necessary to holistically explain multi-wavelength afterglow emission \citep[e.g.][]{Huang2016,Laskar2019}. Our single synchrotron component spectral fits provide a reasonable fit to the majority of the observed data points (Figure \ref{sec:spec_modelling}), but in some individual epochs, e.g. at 42 and 169\,d, there is evidence that a single peaked spectrum insufficiently captures the shape of the observed spectrum. This structure in the spectra could be due to multiple emission components emerging. Our spectral fitting favours a forward shock dominating the observed radio emission, however detailed multiwavelength afterglow modelling, beyond the scope of this work (Rohde et al. in prep), is necessary to fully constrain the regime and physics of the shock and the relative reverse and forward shock contributions. 

\subsubsection{Spectral fits including X-ray data}\label{sec:x-ray_fits}
\citet{OConnor2025} presented detailed X-ray observations of GRB 250702B and found the X-ray light curve is consistent with a power-law decay of $t^{-1.7}$ to $t^{-1.9}$. They also modelled the X-ray spectrum of the afterglow and found weak evidence of a cooling break at $\sim10$\,keV at 1.6\,d post-burst, although noted that a single power-law also provided an adequate fit to the data. In the scenario in which the X-ray emission is produced by the synchrotron emission powering the radio emission, we might expect the radio spectrum to extrapolate to the X-ray measurements. Furthermore, a cooling break between the radio to X-ray bands may be detectable based on the spectral shape. Therefore, in this section we re-do the spectral fits but include an extrapolation of the 0.3--10\,keV X-ray flux density measurements for each epoch. To extrapolate the X-ray flux densities, we use the best-fit power-law model from \citet{OConnor2025} for the \textit{Swift}-XRT data at 0.3--10\,keV assuming $t_0$ is the GRB trigger time and excluding the XRT data prior to 0.55\,d due to the high amplitude variability at early times. The resulting function is $F_X = 2.07\times10^{-11}t^{-1.79}$\,erg\,cm$^{-2}$\,s$^{-1}$. For the spectral fitting we assume a conservative error of 30$\%$ for the X-ray flux measurements. 

\begin{figure*}
    \centering
    \includegraphics[width=\linewidth]{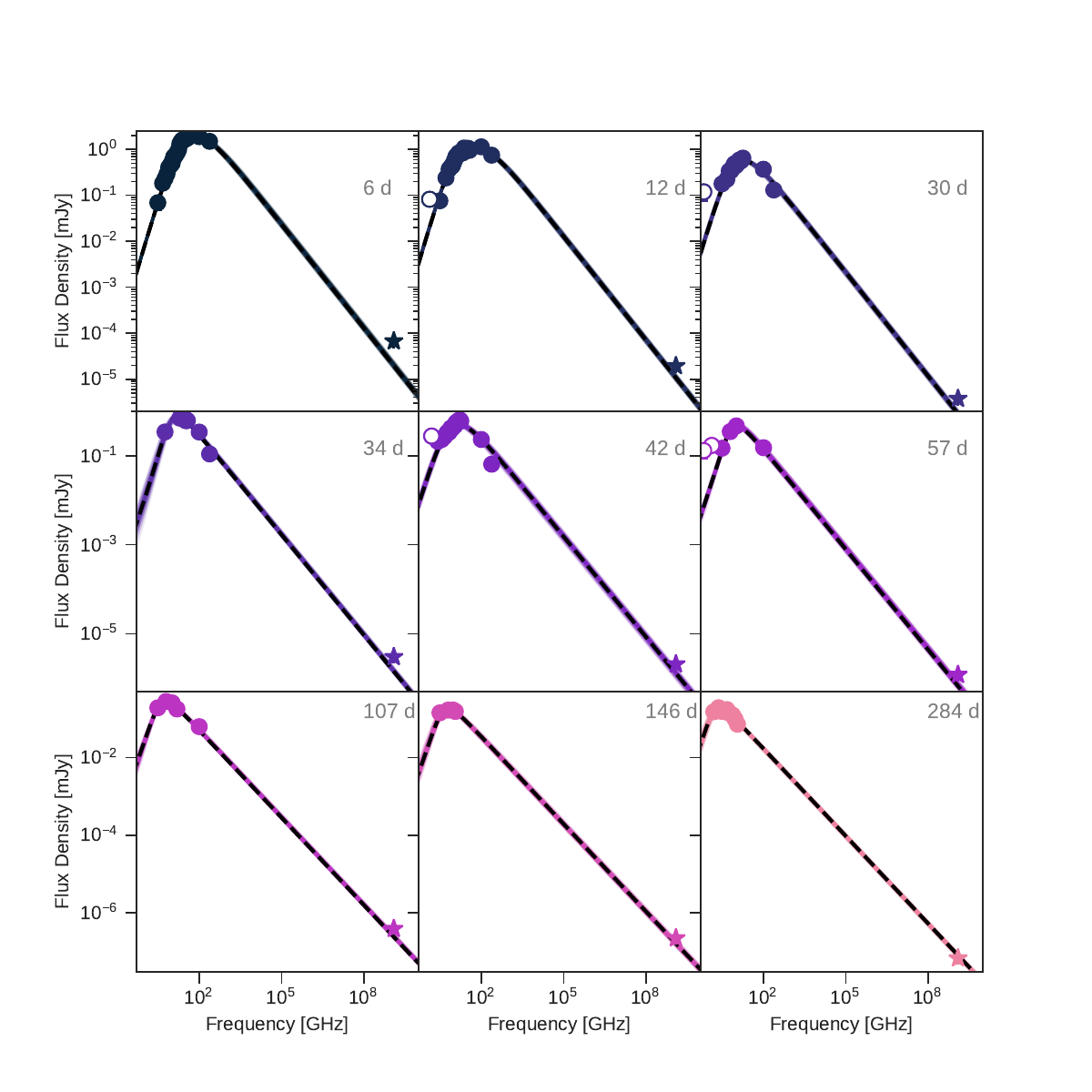}
    \caption{Spectral fits including the radio+X-ray measurements with the synchrotron electron index $p$ fixed to $p=2.5$ in both cases. The coloured lines show the spectral fits for the radio+X-ray data whereas the dashed black lines show the spectral fits for just the radio data. Overall the spectrum all the way up to X-ray frequencies is well fit by the synchrotron model, except at early times (6--42\,d), where there is an excess of X-ray emission above the synchrotron model. Note the many orders of magnitude that both axes span.}
    \label{fig:x-rayfits}
\end{figure*}

We fit the radio+X-ray spectra in the same manner as outlined in Section \ref{sec:spec_modelling}, including fixing $p=2.5$. We show the spectral fits including the X-ray flux measurements in Figure \ref{fig:x-rayfits} (coloured lines), compared to an extrapolation of the radio-only spectral fits (with fixed $p$, dashed black lines). We find the synchrotron model fits the radio and X-ray data well for each epoch, except at early times where there is an excess of X-ray emission. We find no requirement for the cooling break to be present between the radio and X-ray bands. When interpreting Figure \ref{fig:x-rayfits} it is important to keep in mind that the frequency and flux density ranges covered span more than 6 orders of magnitude. The early time ($\lesssim34$\,d observer frame) X-ray excess is a significant deviation from the model, and implies either the simple broken power-law spectral model insufficiently describes the synchrotron spectrum of the source or that there is a different physical process powering (some of) the X-ray emission.

\subsubsection{Cooling break}\label{sec:cool_break}

In the scenario in which sufficiently broad frequency coverage is available, a cooling break may be detected in the regime where  $\nu_m < \nu_a < \nu_c$ or $\nu_a < \nu_m < \nu_c$ and Equations \ref{eq:Fv2} and \ref{eq:Fv1} may be multiplied by 

\begin{equation}\label{eq:cool_break}
\left[1 + \left(\frac{\nu}{\nu_{\rm c}}\right)^{s_3(\beta_6 - \beta_7)}\right]^{-1/s_5}
\end{equation}
where $\beta_6 = \frac{1-p}{2}$, $\beta_7 = -\frac{p}{2}$, and $s_5$ is a smoothing parameter \citep{Granot2002} which may be steeper than the theoretical $s_5=0.8 - 0.03p$ \citep[e.g.][]{Eftekhari2018, Cendes2021}; in our fits we use $s_5=10$ as in \citet{Cendes2021} and \citet{Eftekhari2018}. 

\begin{figure*}
    \centering
    \includegraphics[width=\linewidth]{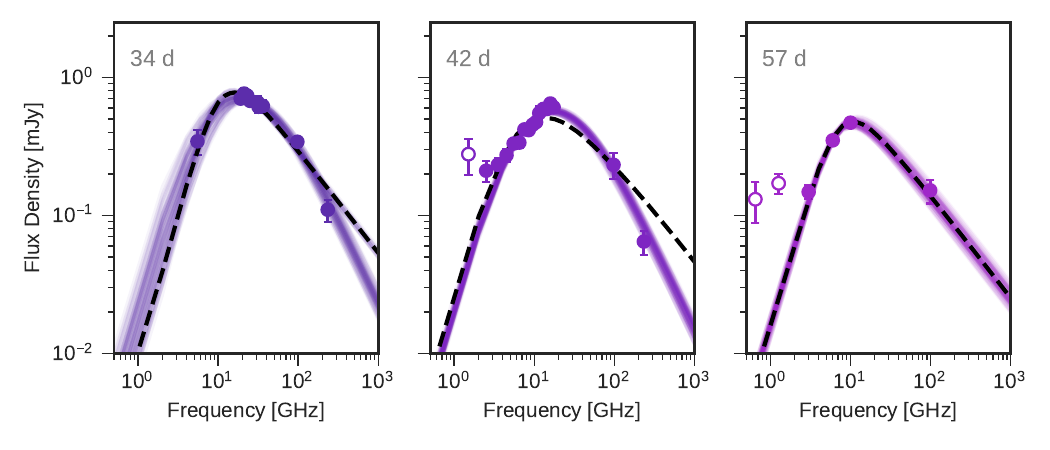}
    \caption{Spectral fits at 34, 42, and 57\,d including an additional spectral component for the cooling break $\nu_c$ (coloured lines). The spectral fits with no cooling break are plotted as dashed black lines for reference, showing including a cooling break provides a better fit to the high frequency ALMA data at 42\,d (with $\nu_c=62\pm30$\,GHz), but no constraint on the cooling break is obtained at 34 and 57\,d.}
    \label{fig:spec_fits_cool}
\end{figure*}

The spectral epoch taken at 42\,d post-burst (plotted in Figure \ref{fig:spec_fits}) shows evidence  of a steepening at the high frequency end of the spectrum that is not well-modelled by the spectral model including only the $\nu_m$ and $\nu_a$ spectral breaks. We therefore fit the 42\,d epoch and the two epochs either side (34 and 57\,d) including the additional spectral component in Equation \ref{eq:cool_break}. We use the same MCMC approach outlined in Section \ref{sec:spectral_fitting}, with the additional fit parameter $\nu_c$, allowing $1 < \log_{10}\nu_c \rm{[GHz]} < 10$. The resulting spectral fits are plotted in Figure \ref{fig:spec_fits_cool}. $\nu_c$ was only constrained by the fit for the 42\,d epoch, with no constraint on $\nu_c$ obtained at 34 and 57\,d. The spectral fit including $\nu_c$ results in fit values of $\nu_m$ and $\nu_a$ which agree within uncertainty of the fit without $\nu_c$. Fitting all other epochs in time with a cooling break does not provide a constraint on the cooling break. Therefore, given we find only one epoch in which including a cooling break provides a better fit to the data, and given the X-ray flux measurements shown in Section \ref{sec:x-ray_fits} do not require a cooling break for the observed spectral index between the radio and X-ray measurements, we deduce that the cooling break is likely above the radio band in all observations. A high frequency for the cooling break is in keeping with the results of \citet{OConnor2025} who found tentative evidence for a cooling break in the X-ray data at early times. The singular evidence of the cooling break at 42\,d in our radio observations is likely a statistical anomaly, or evidence of some decorrelation in the 233\,GHz ALMA data, which is the one data point that requires the cooling break. 

\section{Equipartition analysis}\label{sec:equi}

Next, using the synchrotron spectral properties fit above, we estimate the physical properties of the outflow at each spectral epoch under the assumption of equipartition. While numerous assumptions are made in the equipartition method, the approach allows for a robust measurement of the minimum energy in the outflow and can provide order-of-magnitude constraints on the physical properties of the blastwave as well as the local environment. Whilst in this work we use the equipartition derivations from \citet{BarniolDuran2013} as an order of magnitude estimate, more precise correction factors and considerations are discussed in \citet{Rohde2026}, which do not change the order of magnitude estimates we present here. 

We use the equipartition derivations of \citet{BarniolDuran2013} for a relativistically moving synchrotron source, assuming an on-axis relativistic jet. The radius is given by 

\begin{equation}
\begin{aligned}\label{eq:equi1}
    R = 1.7\times10^{17} \left(\frac{F_p}{\rm{mJy}}\right)^{8/17} \left(\frac{d_L}{10^{28}\rm{cm}}\right)^{16/17} \left(\frac{\nu_p}{10\rm{GHz}}\right)^{-1} \eta^{35/51} \\
    (1+z)^{-25/17} \Gamma^{10/17} f_A^{-7/17} f_V^{{-1/17}} \\
    \chi_e^{(2-p)/34}\xi^{1/17} \epsilon^{1/12}\,\, [\rm{cm}]
\end{aligned}
\end{equation}
where $F_p$ is the peak flux density of the synchrotron spectrum, $\nu_p$ the peak frequency, $d_L$ the luminosity distance (for GRB 250702B at $z=1.036$ we use $d_L=2.192\times10^{28}$\,cm), $\eta=\nu_m/\nu_a$ for $\nu_a<\nu_m$, or $\eta=1$ for $\nu_a > \nu_m$, $f_A$ and $f_V$ are geometric factors where $f_A = \frac{A}{\pi R^2/\Gamma^2}$ and $f_V = \frac{V}{\pi R^3/\Gamma^4}$ for an outflow with area, $A$, and volume $V$, and $\Gamma$ is the bulk Lorentz factor. For narrow ($\theta_j<1/\Gamma$) jet geometries we use the narrow geometry defined in \citet{BarniolDuran2013} where $f_A=f_V=f_{\theta}=(\theta_j\Gamma)^2$. We include the additional correction factors $\chi_e=1$ ($\nu_a < \nu_m$) or $\chi_e=\nu_m/\nu_a$ ($\nu_a > \nu_m$) for electrons that radiate at $\nu_m$, $\epsilon = (\epsilon_B/\epsilon_e)(11/6)$ accounts for deviation from equipartition where $\epsilon_B$ and $\epsilon_e$ are the fraction of the protons energy that go into the magnetic field and the electrons respectively, and $\xi=1 + (1/\epsilon_e)$ is the hot proton correction.

$\Gamma$ can be calculated if the launch date of the relativistic outflow is known by assuming $t\approx R(1+z)/(2c\Gamma^2)$, which is standard special relativity when $\Gamma \gg 1$, and therefore

\begin{equation}
\begin{aligned}
    \Gamma = 12  \left(\frac{F_p}{\rm{mJy}}\right)^{1/3} \left(\frac{d_L}{10^{28}\rm{cm}}\right)^{2/3} \left(\frac{\nu_p}{10\rm{GHz}}\right)^{-17/24} \eta^{35/72} \\ (1+z)^{-1/3} t_d^{-17/24}  f_A^{-7/24} f_V^{{-1/24}} \epsilon^{1/24}
\end{aligned}
\end{equation} 

where $t_d$ is the time since outflow launch in days. 

The equipartition energy is given by

\begin{equation}
    \begin{aligned}
        E = 2.5\times10^{49} \left(\frac{F_p}{\rm{mJy}}\right)^{20/17} \left(\frac{d_L}{10^{28}\rm{cm}}\right)^{40/17} \left(\frac{\nu_p}{10\rm{GHz}}\right)^{-1}\\ \eta^{15/17}
    (1+z)^{-37/17} \Gamma^{-26/17} f_A^{-9/17} f_V^{{6/17}}
     \chi_e^{11(2-p)/34}\\ \xi^{11/17} \left(\frac{11}{17}\epsilon^{-5/12} + \frac{6}{17}\epsilon^{7/12}\right)\,\, [\rm{erg}].
    \end{aligned}
\end{equation}

We also provide estimates of the outflow ambient density ($n_e$), magnetic field ($B$), and velocity $\beta = v/c$ using equations 15, 16, and 22 from \citet{BarniolDuran2013}.

\begin{figure*}
    \centering
    \includegraphics[width=\linewidth]{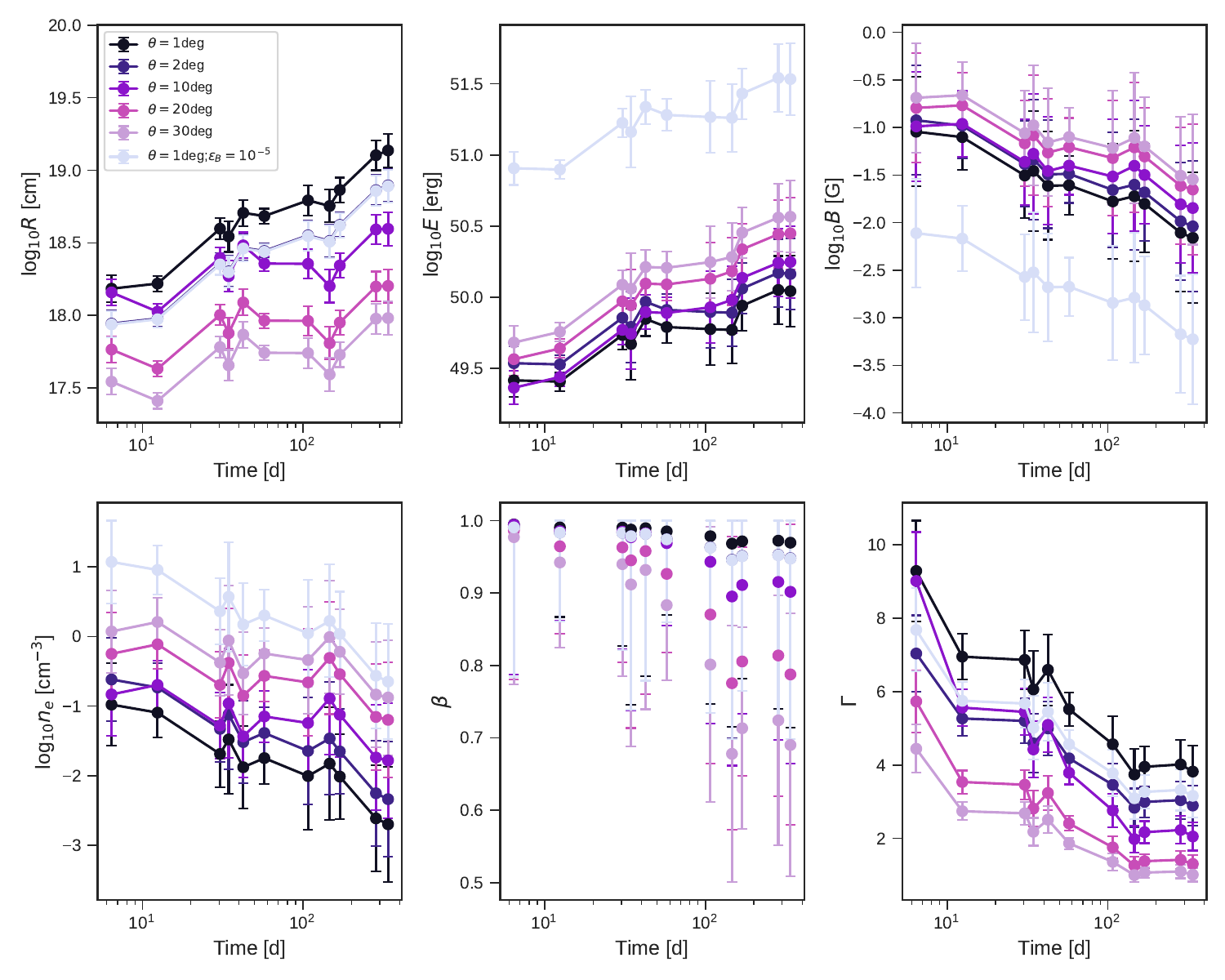}
    \caption{Inferred parameters of the jet and its environment from equipartition. We show six different sets of model parameters for different assumed jet geometries and shock microphysics. Where unspecified, we assume fiducial values of $\epsilon_e=0.1$ and $\epsilon_B=0.01$}
    \label{fig:equi_analysis}
\end{figure*}

In Figure \ref{fig:equi_analysis} we plot the resulting inferred physical properties of the jet and its environment for six different assumptions: a wide, $30$\,deg, $20$\,deg, and $10$\,deg angle outflow with $\epsilon_e=0.1$ and $\epsilon_B=0.01$, a narrow jet with $\theta_j=1$\,deg, or $2$\,deg with $\epsilon_e=0.1$ and $\epsilon_B=0.01$, and a narrow jet with $\theta_j=1$\,deg, $\epsilon_e=0.1$ and $\epsilon_B=10^{-5}$. The 1\,deg, $\epsilon_B=10^{-5}$ case is selected based on the afterglow fit results of \citet{OConnor2025,Granot2026}. In all cases, the shock is initially relativistic, with $\beta\approx1$ at 6\,d and $\Gamma>5$. We see the radius increasing from $\sim10^{17}-10^{19}$\,cm, energy increasing from $\sim10^{49.5}-10^{51}$\,erg, and ambient density decreasing from $\sim10^{0.5}-10^{-2}$\,cm$^{-3}$ over the course of the observations. For a moderately collimated jet, the equipartition analysis requires a low Lorentz factor of $\approx$5 at 6\,d (observer frame), decreasing to $\approx$1 around 100\,d post-burst (observer frame). Even for the extremely narrow jet model, the Lorentz factor is $\approx10$ at 6\,d (observer frame), requiring significant deceleration by 6\,d (observer frame) relative to the inferred values of $\Gamma_0$ from previous afterglow modelling in \citet{OConnor2025} and \citet{Granot2026}, who found $\Gamma_0\gtrsim80$. 

In Section \ref{sec:cool_break}, we found it likely that the synchrotron cooling break lies above the X-ray band. A measurement of the cooling break provides an additional equation to solve for the physical outflow properties, allowing a constraint on $\epsilon_B$ to be obtained due to the dependence of the synchrotron cooling break on the magnetic field. \citet{Sari1998} define the cooling break 

\begin{equation}\label{eq:cool}
    \nu_c = \frac{q_e B \Gamma \gamma_c^2}{2\pi m_e c}
\end{equation}
where 

\begin{equation}
    \gamma_c = \frac{6 \pi m_e c}{\sigma_T B^2 \Gamma t}
\end{equation}
and $q_e$ is the electron charge, $m_e$ the electron mass, $\sigma_T$ the Thomson cross section, and $t$ time since the jet launch.

In the scenario in which the cooling break is located above the X-ray band, we can use this constraint and Equation \ref{eq:cool} to determine an upper limit on $\epsilon_B$. For the $1$\,deg jet geometry, assuming $\nu_c\gtrsim1.2\times10^{18}$\,Hz, we find self-consistent solutions to the equipartition equations if $\epsilon_B\lesssim5\times10^{-6}$. The $30$\,deg jet geometry requires $\epsilon_B\lesssim2\times10^{-7}$. 

The requirement for the cooling break to be above the X-ray band implies extreme deviation from equipartition. Such low values of $\epsilon_B$ are extreme among the GRB population and TDEs where $\epsilon_B$ has been constrained, with most observed GRBs lying in the range $10^{-5}$--0.5 \citep[e.g.][]{Wang2018}, and TDEs $>0.01$ \citep{Cendes2021,Christy2024}. Given the X-ray fluxes used in the radio-X-ray fits in Appendix \ref{sec:x-ray_fits} are extrapolated well beyond the observed time range of the X-rays and a single frequency point is used to fit over $\sim$8 orders of magnitude in frequency, we caution that we cannot rule out a broad cooling break that lies below the X-ray band. Futhermore, the X-ray emission is in excess of our simple synchrotron model at early times, which could indicate a different physical mechanism contributing to the X-ray emission, as was observed for example in the relativistic jetted TDE Swift J1644 \citep[e.g.][]{Eftekhari2018}. The synchrotron X-ray flux could viably be significantly lower than the extrapolated observed lightcurve, and a cooling break between the radio and X-ray bands could be accommodated. Either of these scenarios would alleviate the requirement for such extreme values of $\epsilon_B$. 

\subsection{Consistency with source size estimates from ISS}

\begin{figure}
    \centering
    \includegraphics[width=\linewidth]{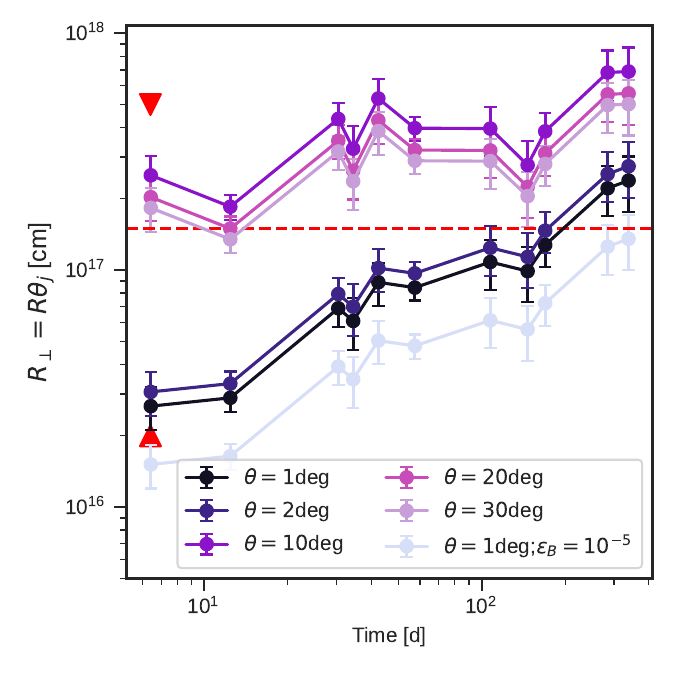}
    \caption{The estimated blastwave image size, $R_{\perp}$, for the different models considered in the equipartition analysis compared to the upper and lower limits on the blastwave image size required by the detection of ISS at early times (red inverted and downward triangles, respectively). The dashed red line shows the approximate average radius of the blastwave image obtained from the ISS induced variability detected at 3\,GHz.}
    \label{fig:Rtheta}
\end{figure}

In Section \ref{sec:disc_flux_excess}, we found evidence of interstellar scintillation causing variability of the low-frequency radio emission, imposing a limit on the blast-wave image size of $1.2\times10^{16}\lesssim R_{\perp} \lesssim 5\times10^{17}$\,cm at early times. While the equipartition analysis robustly constrains the \textit{minimum} source radius, deviations from equipartition have only a small effect on the source radius (with larger effect on the energy). In order to compare with the equipartition radius estimates, we first need to convert the equipartition radius to an approximate blastwave image size. The radius of the observed blastwave image is $R_{\perp}=R\theta_j$ \citep{Granot2005}. In Figure \ref{fig:Rtheta} we plot the estimated blastwave image radius for the various geometries considered in the equipartition analysis. 

The lower limit on the source size is marginally inconsistent with the blastwave image radius for the most extreme model considered ($\theta_j=1$\,deg, $\epsilon_B=10^{-5})$. All other model radii are approximately consistent with the ISS source size constraints. We further discuss the jet geometry and energetics in Section \ref{sec:discussion}. 

\section{Discussion}\label{sec:discussion}

Our radio observations of GRB 250702B reveal evolving synchrotron emission consistent with an initially relativistic outflow likely in the slow-cooling adiabatic expansion phase. The combined temporal and spectral evolution favours a stratified ambient medium, and the observed radio light-curve evolution is broadly consistent with the passage of $\nu_m$ and $\nu_a$ through the observing bands for an adiabatic self-similar forward shock solution. Minor deviations from this model are likely due to jet structure and/or a reverse shock contribution, which future mutliwavelength afterglow modelling efforts may be able to constrain. In this section we explore further the jet properties and potential ejection mechanism, compare the radio emission observed from GRB 250702B with other radio-bright energetic transients, and discuss the possible progenitor system that could produce the properties of the radio jet we have observed.

\subsection{The Lorentz factor and geometry of the jet}

The equipartition analysis employing fiducial values of $\epsilon_e=0.1$ and $\epsilon_B=0.01$ requires moderate bulk Lorentz factors of the source at 6\,d (observer frame, Figure \ref{fig:equi_analysis}). The scintillation observed at low-frequencies at early times requires a blast wave image size $1.2\times10^{16}\lesssim R_{\perp} \lesssim 5\times10^{17}$\,cm. All models except the most extreme narrow jet geometry/far from equipartition satisfy this condition (Figure \ref{fig:Rtheta}). 

A detection of a jet-break in the afterglow emission from the GRB can provide a strong constraint on the jet opening angle \citep[e.g.][]{Sari1999a,Frail2001,Wang2018}. The jet break time ($t_j$) is related to the jet half-opening angle and isotropic energy for a wind medium via \citep{Chevalier2000a,Bloom2003}

\begin{equation}\label{eq:thetaj}
\theta_j = 0.2 \left[\frac{t_{j,\rm{d}} A_{*,\rm{cm^{-3}}}}{(1+z)}\right]^{1/4} \left[\frac{ E_{\rm{iso,\gamma}}\eta_{\gamma}}{10^{52}\rm{erg}}\right]^{-1/4} \rm{rad}
\end{equation}
where $A_*$ is defined as $n = A_*R^{-2}$ for the wind case, $t_j$ is the jet break time in the observer frame, and $\eta_{\gamma}$ is the efficiency of the fireball in converting the energy in the ejecta into gamma-rays.

There is no evidence of a jet break in our radio observations spanning 6--356\,d observer frame post-burst (Figure \ref{fig:radio_lightcurves_fitparams}). There was also no evidence of a jet break in the early-time ($>0.5$\,d) X-ray afterglow observations \citep{OConnor2025}, with a relatively steep X-ray decay, potentially indicating an early jet break prior to the start of the multiwavelength observations. We can therefore deduce that either the jet break occurred before the commencement of the multiwavelength follow-up observations ($t_j<0.5$\,d), or the jet break is yet to occur ($t_j>355$\,d), both reported in the observer frame. 
Assuming $E_{\rm{iso,\gamma}}\approx2.2\times10^{54}$\,erg \citep{Neights2026}, $\eta_{\gamma}\approx0.2$, and $A_*\approx1$\,cm$^{-3}$, with Equation \ref{eq:thetaj} we find $\theta_j\lesssim2$\,deg (early jet break) or $\theta_j\gtrsim15$\,deg (late jet break).

The early jet break scenario is consistent with the opening angle inferred from multiwavelength modelling of the early time data of $\theta_j\lesssim1$\,deg \citep{OConnor2025,Granot2026}, although these initial afterglow models provided a relatively poor fit to the early low-frequency radio data. A jet break beyond the time span of our radio observations (355\,d) would require a larger opening angle of $\theta_j\gtrsim15$\,deg. Both of these scenarios produce sufficiently compact radio-emitting region size at early times to be consistent with the observed ISS based on our equipartition analysis (Figure \ref{fig:equi_analysis}). The temporal evolution of the synchrotron spectral properties are smooth and most consistent with a self-similar adiabatically expanding forward shock. These properties favour a wider low-Lorentz factor jet with no strong reverse shock component in the radio. However, if the cooling break is above the X-ray band and the X-ray emission also arises in the same forward shock, this solution would require even more extreme deviation from equipartition $\epsilon_B\lesssim2\times10^{-7}$. Such a low value of $\epsilon_B$ is extremely unusual in observed synchrotron shocks \citep[e.g.][]{BarniolDuran2013}. It is possible the X-ray emission in excess of the extrapolation of the radio synchrotron spectra (Figure \ref{fig:x-rayfits}) is dominated by a different component, for example a reverse shock or other physical process, which would alleviate the requirements for such a low value of $\epsilon_B$. 
We therefore deduce either a narrow jet or a wider-angle outflow are viable possibilities for the jet opening angle.

The equipartition analysis also provides constraints on the bulk Lorentz factor, $\Gamma$. In all models considered, $\Gamma\lesssim10$ in our first epoch of radio observations at 6\,d (observer frame), even for the narrow $\theta_j=1$\,deg model. The $\sim10$\,MeV Gamma-ray emission detected from this source at early times requires $\Gamma_{0,\rm{min}}\sim50$ \citep{Neights2026}, which would require significant deceleration of the jet in the first 6\,d (observer frame), which is not unusual for GRB-like jets. 
An alternate possibility is that the early-time Gamma-ray and X-ray emission is produced by a separate component than the late-time radio afterglow. In this scenario, a two-component structured jet consisting of a faster relativistic component ($\Gamma \gtrsim 10$) may drive the initial high-energy emission, and a slower, energy-dominated outflow may drive the long-lived forward shock ($\Gamma \lesssim 10$) producing the radio emission. The observed X-ray properties are broadly consistent with a structured outflow in which a narrow, ultra-relativistic jet produces the prompt gamma-ray and early X-ray emission (which we found to be in excess of the radio synchrotron emission, see Appendix \ref{sec:p_fits}), while a slower, energy-dominated outflow drives the forward shock responsible for the late-time afterglow. Such a jet configuration is not typical of standard GRB jets, in which the bulk of the energy is often assumed to reside in the narrow, relativistic core, and departs from the simplest top-hat jet models \citep[e.g.][]{Piran2004, Kumar2015}.

There are therefore two viable scenarios for the jet geometry and bulk Lorentz factor: a narrow $\theta_j\lesssim2$\,deg , high $\Gamma_0$ jet with significant early-time deceleration, or a wider $\theta_j\gtrsim15$\,deg, lower $\Gamma_0$ jet with continuous deceleration over the course of the observations. Future work that incorporates the full multiwavelength emission in detailed afterglow modelling may be able to distinguish between these two scenarios (Rhode et al. in prep), although initial afterglow modelling efforts show it is challenging to self-consistently reproduce the radio and high energy emission with a single evolving synchrotron component \citep[e.g.][]{Granot2026}. 

The narrowly collimated jet scenario provides an opportunity to estimate the total beaming corrected energy of the relativistic jet. The isotropic-equivalent gamma-ray energy was constrained to be $E_{\rm{iso}}\gtrsim2.2\times10^{54}$\,erg \citep{Neights2026}, however, the true energy of the relativistic jet depends on its opening angle. For a narrowly collimated jet with $\theta\sim1$\,deg \citep[e.g.][]{Granot2026,OConnor2025}, the beaming factor is $f_b\approx \theta^2/2$. The true energy of the jet is then $E_{\rm{jet}}\sim f_b E_{\rm{iso,\gamma}} / \eta_{\gamma} \approx 3\times10^{51}$\,erg for $\eta_{\gamma}=0.1$. This beaming corrected energy is entirely consistent with the observed GRB population \citep{Frail2001,Wang2018}, although such a narrow jet opening angle is uncommon among GRB populations \citep{Frail2001,Goldstein2016}.

\subsection{The density of the ambient environment}

The radio emission from a relativistic outflow is strongly dependent on the ambient environment it propagates through \citep[e.g.][]{DeColle2012}. The environment density and stratification can provide key insight into the nature of the transient progenitor as different progenitors are likely to inhabit significantly different environments. An ultra-long GRB from a collapsar interpretation of GRB 250702B would most likely occur in a stellar-wind environment ($k=2$) due to the massive star progenitor. Whereas, the environment expected for a TDE interpretation of the transient is less well-defined, especially in the case of an IMBH-TDE due to the uncertainty in their formation and evolution. However, a black hole may be expected to influence its environment via Bondi (spherical) accretion, with an expected density stratification of $k=1.5$ \citep[e.g.][]{Granot2026,Goodwin2026}.  

\begin{figure}
    \centering
    \includegraphics[width=\linewidth]{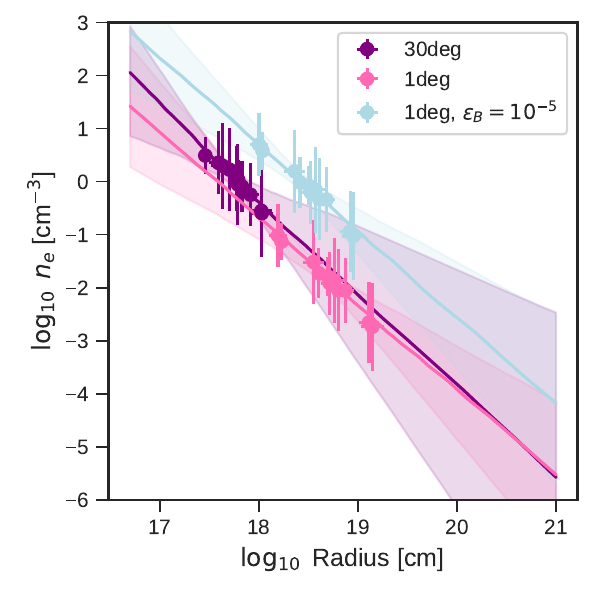}
    \caption{The inferred ambient density with radius based on the equipartition modelling (circles), with a power-law fit (lines) for a narrow jet (pink) and standard jet (purple). A simple power-law fit to these data with an equation of the form $n_e = n_0 (\frac{R}{10^{16}\rm{cm}})^{-k}$ finds $\log_{10}n_0 = 3\pm1; 2.6\pm0.9; 3.9\pm0.9$\,cm$^{-3}$ and $k=1.7\pm0.6; 1.6\pm0.4; 1.6\pm0.4$ for the 30\,deg, 1\,deg, and 1\,deg $\epsilon_B=10^{-5}$ models respectively. In all cases, the ambient density distribution traces $k=1.6$, consistent with a stellar wind environment ($k=2$) or Bondi profile ($k=1.5$), but inconsistent with a flat $k=0$ interstellar medium environment.}
    \label{fig:ambient_density}
\end{figure}

In Figure \ref{fig:ambient_density} we plot the equipartition inferred ambient density with radius for different assumed jet geometries and equipartition fractions. A simple power-law fit to these data with an equation of the form $n_e = n_0 \left(\frac{R}{10^{16}\rm{cm}}\right)^{-k}$ finds $\log_{10}n_0 = 3\pm1; 2.6\pm0.9; 3.9\pm0.9$\,cm$^{-3}$ and $k=1.7\pm0.6; 1.6\pm0.4; 1.6\pm0.4$ for the 30\,deg, 1\,deg, and 1\,deg $\epsilon_B=10^{-5}$ models respectively. 

These fits rule out a flat $k=0$ ISM-like environment to high confidence, consistent with the conclusions of the temporal evolution of the critical synchrotron frequencies, analysed in Section \ref{sec:spectral_fitting}. The ambient density fits are consistent with both a stellar-wind or Bondi profile. These values of $k$ from our equipartition analysis are also consistent with those required by the early-times afterglow modelling \citep{OConnor2025,Granot2026}.  
We therefore deduce that the inferred ambient density gradient lies somewhere between a steady stellar wind or simple Bondi sphere. This finding could suggest an interesting physical story about the progenitor system, although, it is not unusual for some individual supernovae and TDE systems to have environments that deviate from the simplest expectations \citep[e.g.][]{Wellons2012,Alexander2020}.

\subsection{Comparison to GRB and TDE populations}
\begin{figure*}
    \centering
    \includegraphics[width=0.49\linewidth]{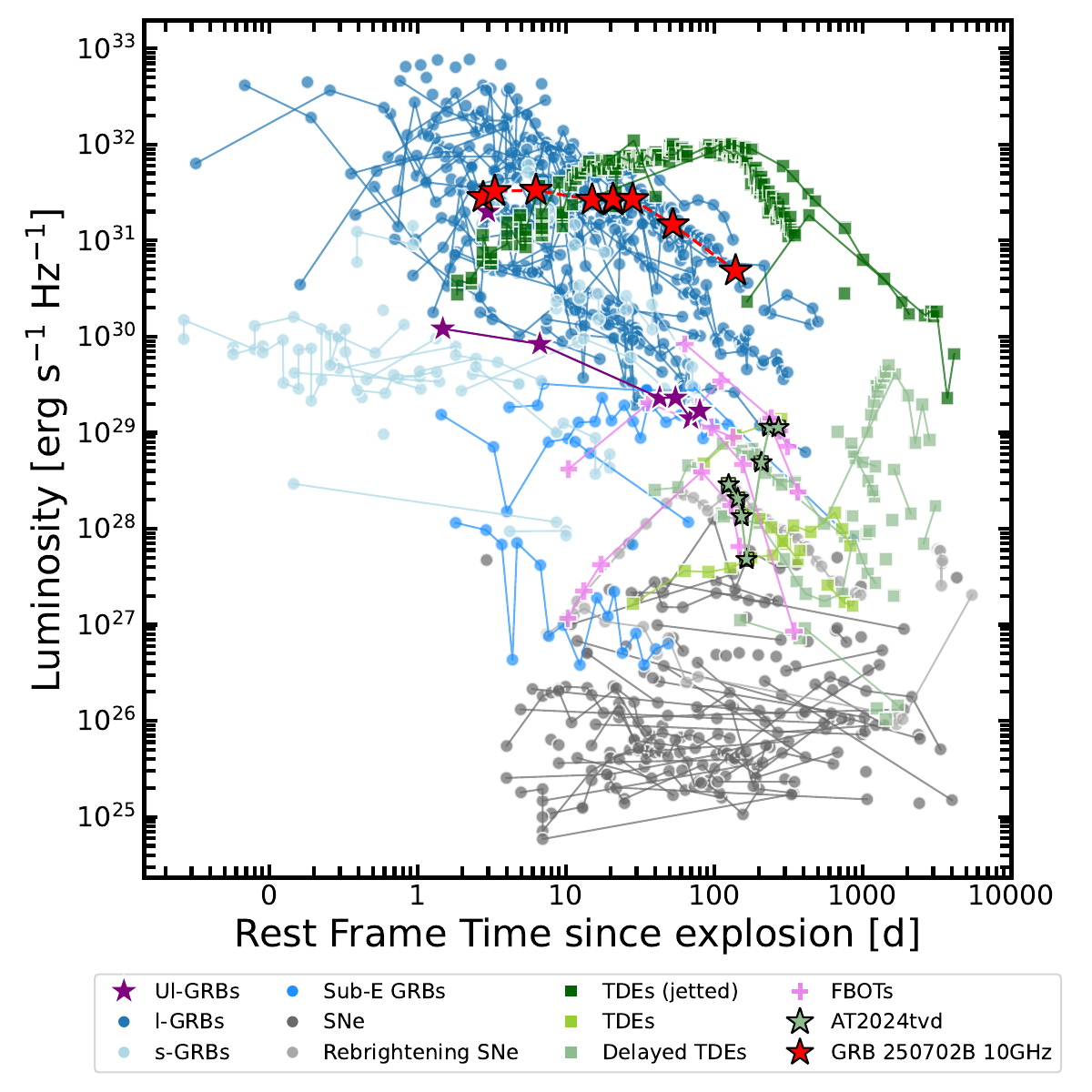}
    \includegraphics[width=0.49\linewidth]{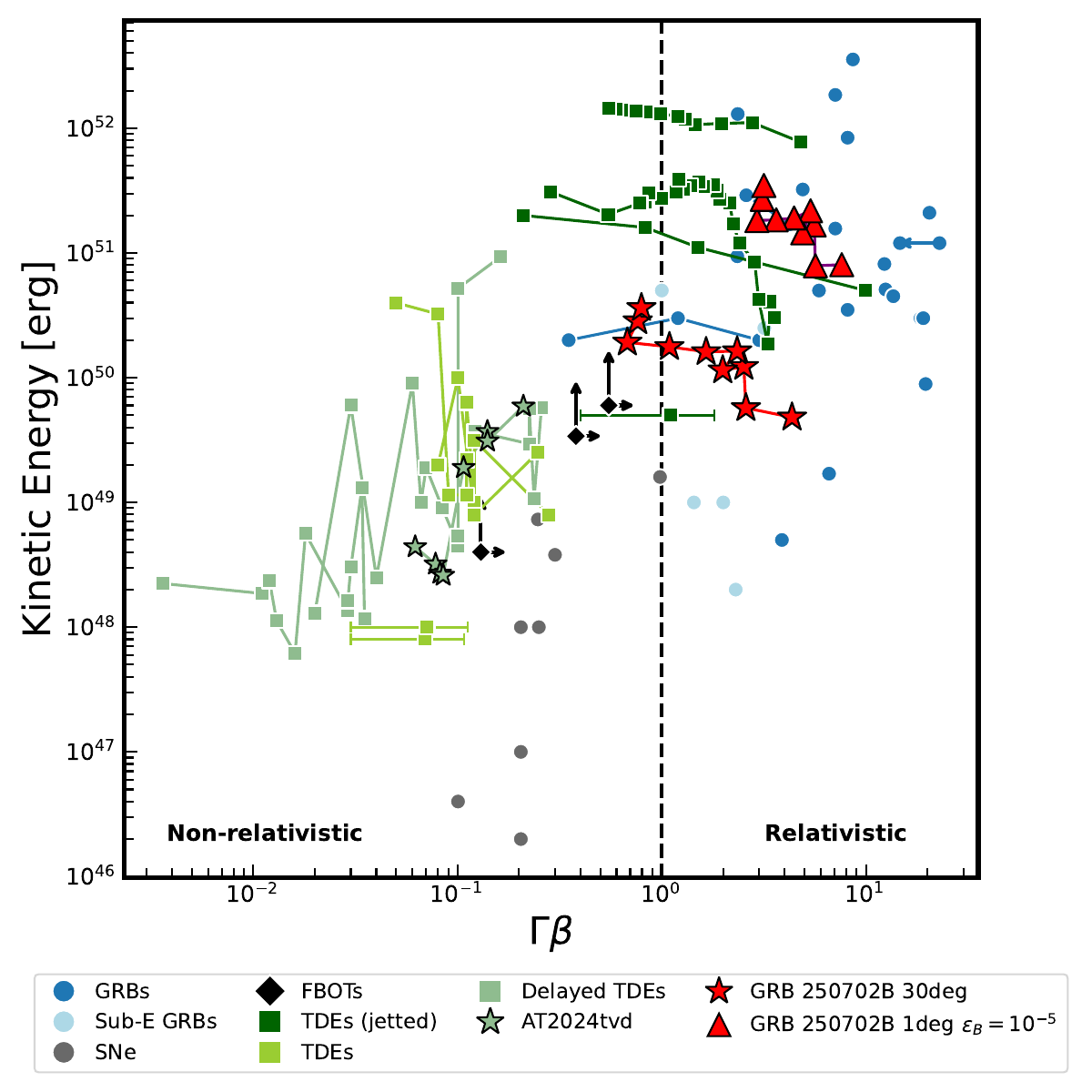}
    \caption{\textit{Left:} The observed luminosity at $\approx$10\,GHz for GRB 250702B (red stars) compared to a selection of other types of energetic transients. \textit{Right:} The inferred outflow kinetic energy from equipartition analysis plotted against the Bulk Lorentz factor, $\Gamma$, multiplied by the outflow velocity, $\beta$ for GRB 250702B (red stars) and a selection of other transients. The off-nuclear TDE AT2024tvd is highlighted in black outlined green stars in both panels for comparison. Plots are adapted from \citet{2026ApJ..1000..118G}. The individual datasets are from \citet{2012ApJ...746..156C,alexander_discovery_2016,alexander_radio_2017,andreoni_very_2022,berger_grb_2001,berger_host_2001,berger_jet_2000,berger_radio_2003,bietenholz_radio_2021,bright_radio_2022,brown_latetime_2017,cendes_mildly_2022,cendes_radio_2021,cendes_ubiquitous_2024,cenko_afterglow_2011,cenko_multiwavelength_2006,cenko_swift_2012,chandra_comprehensive_2008,chandra_discovery_2010,chrimes_multiwavelength_2024,coppejans_mildly_2020,djorgovski_afterglow_2001,eftekhari_associating_2018,frail_accurate_2005,frail_energetic_2006,frail_enigmatic_2000,frail_radio_1999,galama_bright_2000,galama_continued_2003,goodwin_radio_2023,goodwin_systematic_2025,greiner_unusual_2013,harrison_broadband_2001,harrison_optical_1999,ho_at2018cow_2019,ho_koala_2020,horesh_are_2021,horesh_delayed_2021,horesh_unusual_2015,laskar_radio_2023,laskar_reverse_2016,laskar_vla_2018,2014ApJ...780..118F,2022ApJ...935L..11L,2019ApJ...883...48L,2021ApJ...906..127F,2024ApJ...975L..13A,2006ApJ...650..261S,2024Natur.626..737L,2005Natur.438..988B,2021ApJ...906..127F,anderson2025arXiv250814650A,law_discovery_2018,leung_search_2021,margutti_embedded_2019,margutti_signature_2013,mattila_dust-enshrouded_2018,meyer_late-time_2025,moin_radio_2013,mooley_latetime_2022,oconnor_structured_2023,pasham_multiwavelength_2015,perley_afterglow_2014,perley_grb_2008,rhodes_rocking_2024,rol_grb_2007,rose_latetime_2024,schroeder_long-lived_2025,schroeder_radio_2024,sfaradi_offaxis_2024,Sfaradi2025,soderberg_constraints_2004,soderberg_redshift_2004,soderberg_relativistic_2006,stein_tidal_2021,taylor_discovery_1998,van_der_horst_detailed_2008,zauderer_radio_2013,2017ApJ...837..153A,2024A&A...691A.329C,1998Natur.395..663K,1997Natur.389..261F,2025ApJ...983...29H,2012GCN.12804....1H,2023GCN.33475....1A,2008ApJ...688..470P,2002ApJ...572L..51P,2008A&A...480...35V,2026ApJ..1000..118G}}
    \label{fig:grb_comp}
\end{figure*}

The gamma-ray duration of GRB 250702B is unprecedented, but the question remains whether other properties of this source are also exceptional. In this section we compare the observed radio properties of the event with other energetic transients such as GRBs (including ulGRBs) and TDEs. In Figure \ref{fig:grb_comp} we show the radio luminosity evolution at 10\,GHz for a host of different extra-galactic transients. GRB 250702B clearly lies in the high-luminosity region of the plot, with luminosities consistent with standard GRBs and relativistic TDEs. However, the timescale of evolution for relativistic TDEs is typically significantly longer than was observed for GRB 250702B, with the peak and decay rate much more consistent with the standard lGRB population. Interestingly, the other ulGRB plotted shows a peak observed luminosity approximately two orders of magnitude lower than GRB 250702B. 

In Figure \ref{fig:grb_comp} right we plot the estimated kinetic energy from equipartition analysis against the specific momentum (i.e. the bulk Lorentz factor multiplied by $\beta$) for GRB 250702B and a selection of GRBs, SN, FBOTs, and TDEs. The low Lorentz factor and lower kinetic energy for the $\theta_j=30$\,deg case for GRB 250702B is extremely unusual for GRBs, but broadly consistent with the lower energy range of the relativistic TDE population. For the $\theta_j=1$\,deg case the energy and Lorentz factor are broadly consistent with the GRB population.

The most relevant TDE population to compare the observed radio properties of GRB 250702B with is perhaps off-nuclear TDEs, of which there are few known \citep{Guolo2026}. AT2024tvd is the only off-nuclear TDE with extensive radio coverage \citep{Sfaradi2025}, although is inferred to have occurred around a $\sim10^{6}$\,M${_\odot}$ SMBH \citep{Yao2025}. The radio emission observed from AT2024tvd is significantly less luminous than that observed from GRB 250702B, and is unlikely to be produced by an on-axis relativistic jet \citep{Sfaradi2025}, making comparisons between the total energetics and jet/outflow evolution difficult due to the likely different mechanisms powering the radio emission. However, an interesting comparison to explore is the inferred properties of the environment surrounding the off-nuclear black holes. For AT2024tvd, \citet{Sfaradi2025} infer a steep density profile, with $k=3.8$, and $n_0\sim10^{5}$\,cm$^{-3}$, significantly denser and steeper than the environment we infer based on the radio evolution of GRB 250702B. Such different environments may indicate significantly different evolutionary pathways to form the black holes involved in the two events, and suggests the black hole in GRB 250702B is unlikely to be an SMBH, although we note with just one off-nuclear TDE to compare to it is challenging to make wider inferences. Very few bona fide IMBH TDE candidates are known, and even fewer with constrained radio properties. The IMBH TDE candidate AT2020neh did not produce detectable radio emission at early times \citep{Angus2022}, although a faint radio detection was reported at late times \citep{Cendes2024}, suggesting that on-axis relativistic jets are likely also rare among IMBH TDEs. 

\subsection{The nature of the progenitor}\label{sec:progenitors}

The multi-wavelength properties of GRB 250702B taken together with our radio observations provide an opportunity to explore viable progenitor systems. While no single observable is definitive, the combination of outflow energetics, outflow structure, environment, and host-galaxy location taken together allow some insight. There are a number of puzzling observational constraints which any viable model for the progenitor must be able to explain. These include:

\begin{enumerate}
    \item The duration of the gamma-ray emission ($>25,000$\,s) \citep{Neights2026}.
    \item The $\approx1$\,d X-ray precursor detected by EP \citep{Zhang2026}.
    \item The $\sim$10\,MeV emission detected by Fermi \citep{Neights2026}.
    \item The timescale of variability of the gamma-ray and X-ray emission, including flaring episodes in X-ray \citep{OConnor2025}.
    \item The off-nuclear location of the transient and the unusual, dusty host galaxy \citep{Gompertz2026,Sears2026,Levan2025}.
    \item The long-lived radio counterpart broadly consistent with an adiabatically expanding forward shock and consistent with either a highly relativistic, extremely narrow jet, or a wide angle, low Lorentz factor jet dominating the radio emission (this work).
\end{enumerate}

The presence of strong variability and flaring in the X-ray light curve additionally requires either a contribution from ongoing central engine activity or internal shocks within a relativistic jet, superposed on the evolving external shock emission from the ultrarelativistic component of the jet. There is an excess of X-ray emission from our extrapolated synchrotron model (Figure \ref{fig:x-rayfits}) which hints that the X-ray emission may not be dominated by the synchrotron emission from the jet observed at radio frequencies. Although, this observed excess could also be an indication that the simple broken-power law single-component synchrotron model is insufficient. Below we discuss the two main viable progenitor scenarios for GRB 250702B, but note that additional scenarios, such as a neutron star merger with a stellar core \citep{HutchinsonSmith2024} or more exotic supernovae \citep{Nakauchi2013} may be viable for the ulGRB.  

\subsubsection{A stellar collapse engine}
Given that a standard single-star collapsar is ruled out due to the extraordinary duration of the GRB \citep[see the detailed discussion in][]{Neights2026}, the viable stellar collapse engine scenarios remaining necessarily involve an angular momentum contribution from an orbit, such as the helium-star merger model proposed by \citet{Neights2026} and discussed by \citet{Villar2026,Klencki2026}. Helium star mergers have been recently gaining some traction as a potential system that could produce luminous fast blue optical transients \citep[LFBOTS;][]{Villar2026,Klencki2026}, a class of transients potentially related to ulGRBs, although no gamma-ray emission has been associated with an LFBOT to date \citep[with the exception of the unusual ulGRB 111209A for which an SN-like optical counterpart was observed that some authors have argued could be FBOT-like;][]{Greiner2015,Villar2026}. 

A helium star merger provides a natural mechanism for prolonged central engine activity, as the inspiral and subsequent accretion of a compact object can extend over much longer timescales than in classical collapsars \citep{Neights2026}. The wind-like ambient density profile inferred from the radio modelling is also consistent with mass loss from the massive star progenitor system prior to merger (Figure \ref{fig:ambient_density}), although, the ambient medium profile for known LFBOTs deviates significantly from the ambient density profile we infer for GRB 250702B \citep[e.g. AT2024wpp showed a steep density profile with $k\approx3.1$;][]{Nayana2025}. In the helium star merger model, the prompt emission is produced by an ultra-relativistic jet launched during the collapse and accretion of a stripped stellar core, which favours a highly collimated, high-Lorentz-factor outflow. This is consistent with the narrow jet scenario to explain the observed radio evolution of GRB 250702B, which is well described by a single impulsive blast wave with no evidence for ongoing energy injection. The overall energetics and evolutionary timescales of this jet lie well within the expected distribution for standard GRB jets \citep[Figure \ref{fig:grb_comp};][]{Frail2001}. However, \citet{Klencki2026} point out that the beaming-corrected energy for GRB 250702b is consistent with $\lesssim0.1\,M_{\odot}$ of accreted mass for a jet efficiency of $\eta=0.1$, significantly smaller than the expected accreted mass for a He star merger. The alternative viable scenario for radio emission, in which a substantial fraction of the kinetic energy resides in a wider-angle, mildly relativistic component ($\Gamma\lesssim10$), is unable to be reproduced by a single narrow jet viewed off-axis or by simple deceleration of an initially ultra-relativistic flow. \citet{Ramirez2002} showed that relativistic jets propagating through extended stellar envelopes naturally inflate energetic cocoons that shape the angular structure of the emerging outflow. However, such a wide angle jet and the associated lower energy is highly unusual for collapsar-like systems (Figure \ref{fig:grb_comp}).

\subsubsection{A TDE engine}
The observed combination of relativistic gamma-ray emission, prolonged X-ray activity, and a potentially wide-angle, low Lorentz factor jet is broadly consistent with a TDE. Numerous TDE scenarios have been discussed (and debated) in the literature for GRB 250702B, with no apparent consensus on whether the observed timescales of the GRB emission are compatible with TDE timescales. This has been potentially exacerbated by the very low number of relativistic jetted TDEs discovered to date, all of which are inferred to have occurred around (potentially lower mass) SMBHs \citep[e.g.][]{Levan2011,Andreoni2022,Eftekhari2024}. 

\citet{Neights2026} suggested a TDE progenitor was unlikely for GRB 250702B based on the fact that no relativistic TDE to date has been observed to produce MeV gamma-ray emission, the gamma-ray duration of the known relativistic TDEs is more than an order of magnitude longer than GRB 250702B, and that the observed minimum variable timescale of the gamma-ray emission is $>100$ times smaller than invoked previously for a white dwarf-IMBH TDE \citep{krolik_swift_2011}. Others also have shown that the duration of a white dwarf disruption around an IMBH cannot be longer than $\sim15,000$\,s \citep{Neights2026,OConnor2025,Granot2026}. \citet{Eyles2026} argued that chaotic debris streams from multiple stripping episodes could extend the duration of the a white dwarf IMBH TDE engine, and \citet{Sato2026,Yuan2026,Eyles2026} all state the observed emission timescales are compatible with this scenario. However, \citet{Granot2026,Beniamini2026} disfavour a white dwarf disruption due to too short characteristic fallback and viscous timescales as well as difficulties in providing sufficient mass stripped from the white dwarf to power the relativistic jet observed. \citet{Granot2026} instead propose a main-sequence star IMBH TDE as a viable alternative, and \citet{Beniamini2026} suggest a main-sequence star stellar mass BH TDE is also viable. 

It is clear there has been significant debate, and various relevant timescales used to rule in or out different TDE scenarios for this event. The question remains whether a TDE in any of these regimes would be compatible with the observed timescales and energetics of GRB 250702B, particularly in the context of this work where we have observed smoothly evolving radio emission which does not seem to require energy injection and a potential excess of X-ray emission in the first 34\,d post-burst (observer frame).
In this section, we first recap the critical timescales involved in the various TDE scenarios, then assess the compatibility of these timescales with the observed properties of GRB 250702B. 

We provide a full derivation of the relevant timescales for a TDE in Appendix \ref{sec:TDE_timescales}. We consider four candidate disruption scenarios: a stellar mass black hole disrupting a main-sequence star (sBH-MS), an intermediate mass black hole disruption of a main-sequence star (IMBH-MS) or white dwarf (IMBH-WD), and a supermassive black hole disruption of a main sequence star (SMBH-MS).
We summarise the four characteristic timescales for each candidate disruption channel in Table~\ref{tab:TDE_timescales}. It is important to note that because the orbital time at circularisation $\torb$ and the viscous time $\tvisc$ depend only on the mean density of the disrupted star (and on the disc parameters $\alpha$ and $H/R$), they are independent of black hole mass $\Mbh$ (Eqs.~\ref{eq:torb_norm}--\ref{eq:tvisc_norm}). These timescales are therefore identical across the three main-sequence channels, and only $\tfb$ and $t_{\rm SE}$ depend on the black hole mass. The fallback history also depends sensitively on stellar structure and encounter depth \citep{Guillochon2013}, and lower mass black holes naturally produce much shorter fallback timescales than classical SMBH TDEs \citep{Ramirez2009,MacLeod2014}. Furthermore, $\tfb$ is the time at which fallback \emph{begins} and the decay constant of the $t^{-5/3}$ supply (Eq.~\ref{eq:Mdotfb}), not the time at which accretion ceases: the engine is fed for many fallback times. The duration over which the supply rate remains super-Eddington is instead $t_{\rm SE}$ (Eqs.~\ref{eq:tSE_norm},~\ref{eq:tSE_MR}), and for every channel in Table~\ref{tab:TDE_timescales} this exceeds the observed $\gtrsim2.5\times10^{4}$\,s gamma-ray duration by several orders of magnitude. The basic requirement that the central engine remain active throughout the prompt phase (and potentially through the weeks of subsequent X-ray activity) is therefore met by all four channels and does not, on its own, discriminate between them.

\begin{table*}[t]
\centering
\small
\caption{Characteristic TDE timescales for each candidate channel, evaluated from Eqs.~(\ref{eq:tfb})--(\ref{eq:tSE_MR}) at the fiducial parameters indicated ($m_\star=r_\star=\beta=\eta_{0.1}=1$; thick super-Eddington disc with $\alpha_{0.1}=\theta_{0.3}=1$, except where noted). Because $\torb$ and $\tvisc$ are independent of $\Mbh$, they are common to the three main-sequence columns. Note that $t_{\rm SE}\propto\Mbh^{-2/5}$, so the super-Eddington supply lasts \emph{longest} for the lowest-mass black holes; the literal values for low-mass black holes far exceed any plausible jet lifetime and simply indicate that supply duration never limits these channels.}
\label{tab:TDE_timescales}
\renewcommand{\arraystretch}{1.3}
\begin{tabular}{lcccc}
\hline\hline
 & \textbf{sBH--MS} & \textbf{IMBH--MS} & \textbf{IMBH--WD} & \textbf{SMBH--MS} \\
 & ($\Mbh\!\sim\!10\,\Msun$) & ($\Mbh\!\sim\!10^{4}\,\Msun$) & ($\Mbh\!\sim\!10^{4}\,\Msun$) & ($\Mbh\!\sim\!10^{6}\,\Msun$) \\
\hline
$\tfb$
& $\sim10^{4}$\,s ($\sim$3\,hr)
& $\sim3\times10^{5}$\,s ($\sim$4\,d)
& $\sim3\times10^{2}$\,s ($\sim$5\,min)
& $\sim3.5\times10^{6}$\,s ($\sim$40\,d) \\

$\torb$
& $\sim3\times10^{4}$\,s ($\sim$8\,hr)
& $\sim3\times10^{4}$\,s ($\sim$8\,hr)
& $\sim25$\,s
& $\sim3\times10^{4}$\,s ($\sim$8\,hr) \\

$\tvisc$
& $\sim5\times10^{5}$\,s ($\sim$6\,d)
& $\sim5\times10^{5}$\,s ($\sim$6\,d)
& $\sim4\times10^{2}$\,s ($\sim$7\,min)
& $\gtrsim5\times10^{5}$\,s$^{\dagger}$ \\

$t_{\rm SE}$
& $\sim7\times10^{9}$\,s ($\sim10^{2}$\,yr)
& $\sim4\times10^{8}$\,s ($\sim$13\,yr)
& $\sim2.5\times10^{7}$\,s ($\sim$300\,d)
& $\sim7\times10^{7}$\,s ($\sim$2\,yr) \\

$t_{\rm SE}/\tfb$
& $\sim6\times10^{5}$
& $\sim10^{3}$
& $\sim8\times10^{4}$
& $\sim20$ \\

Hierarchy
& $\tfb\!\lesssim\!\torb\!\ll\!\tvisc$
& $\torb\!\ll\!\tfb\!\sim\!\tvisc$
& $\torb\!\ll\!\tvisc\!\sim\!\tfb$
& $\torb\!\ll\!\tvisc\!\ll\!\tfb$ \\

Flow regime
& evolving (heavily smeared)
& evolving (viscously smeared)
& evolving (viscously smeared)
& quasi-steady \\

Compatible?
& Yes
& Yes
& Yes (see text)
& Disfavoured \\
\hline\hline
\end{tabular}

\smallskip
{\footnotesize $^{\dagger}$ Canonical thick-disc value; a near- or sub-Eddington SMBH disc is geometrically thin ($H/R\!\ll\!0.3$), which lengthens $\tvisc$ substantially and gives $\tfb\!\ll\!\tvisc$.}
\end{table*}

We can now compare these timescales with the observed properties of GRB 250702B. $\tfb$ sets when fallback, and hence jet launching, can first commence. For a main-sequence disruption by an intermediate-mass black hole, $\tfb$ is of order days (Eq.~\ref{eq:tfb}), comparable both to the $\sim1$\,d X-ray precursor detected by EP \citep{EP2025GCN} and to the day-long span of the brightest gamma-ray episodes. For a main-sequence disruption by a stellar-mass black hole, $\tfb\sim$\,hours, similarly consistent with the onset and duration of the prompt phase. In both cases $\torb\ll\tvisc\sim\tfb$, so fallback, circularisation, and viscous accretion proceed concurrently. The result is a prolonged but finite-duration, dynamically evolving central engine, able to power the extended X-ray emission and to launch a structured outflow at early times without requiring late-time energy injection into the blast wave (in keeping with the smooth, single-component radio light curves, Section~\ref{sec:lightcurve_fits}). A white-dwarf disruption by an IMBH has a much shorter fallback time, $\tfb\sim$\,minutes \citep[Eq.~\ref{eq:tfb_MR}][]{LawSmith2017,Granot2026}, with the disc dynamical and viscous times also of order seconds to minutes (Eqs.~\ref{eq:torb_MR}--\ref{eq:tvisc_MR}). However, its super-Eddington supply persists for $\sim300$\,d (Eq.~\ref{eq:tSE_MR}), comfortably spanning the weeks-long X-ray activity.

A central observable is the strong short-timescale variability of the prompt gamma-ray and X-ray emission, which includes $\sim$ks flaring superposed on the X-ray light curve \citep{OConnor2025} and sub-second structure in the gamma-rays \citep{Neights2026}. The $\sim$ks variability is far shorter than the global, circularisation-scale timescales of the main-sequence channels ($\torb\sim8$\,hr, $\tvisc\sim6$\,d; Table~\ref{tab:TDE_timescales}), and the sub-second structure is shorter than the global timescales of all channels; such variability therefore cannot be set by the large-scale accretion clock. It instead reflects the dynamical time of the inner accretion flow and jet base, of order the light-crossing time of a few gravitational radii ($\sim$\,few\,$\times r_g/c$), further compressed in the observer frame by relativistic beaming. Because $r_g\propto\Mbh$, this favours the lower-mass engines: for $10$--$10^{4}\,\Msun$ this inner timescale is $\sim10^{-3}$--$1$\,s, accommodating both the ks and sub-second variability, whereas for a $10^{6}\,\Msun$ black hole it is $\sim$\,minutes, and the outer disc evolves only on the $\sim8$\,hr to multi-day timescales of Table~\ref{tab:TDE_timescales}. The short intrinsic timescales of the white-dwarf--IMBH and stellar-mass channels are therefore an asset for reproducing the observed variability, not a liability.

The radio data add a further, independent constraint: the emission is either dominated by a mildly relativistic ($\Gamma\lesssim10$), wide-angle outflow carrying $E_K\sim10^{50}$\,erg, launched impulsively, (Sections~\ref{sec:equi},~\ref{sec:spectral_fitting}), or a narrow $\theta_j\lesssim2$\,deg highly relativistic jet, also launched impulsively. The bulk of the kinetic energy must therefore be deposited early, during the initial engine phase, with the ejecta stratification in both Lorentz factor and angle established essentially at launch. 
Stratification arises most naturally when fallback, circularisation, and viscous accretion overlap, so that the baryon loading and jet properties vary strongly across a brief, rapidly evolving launch phase. This condition is met by all three low-mass channels. In particular, all three low-mass channels have $\tvisc\gtrsim\tfb$. The ``viscous'' blurring of the fallback (the ratio $\tvisc/\tfb\simeq0.14\,M_6^{-1/2}$) is $\approx1$ for both the white-dwarf--IMBH and main-sequence--IMBH cases (Eqs.~\ref{eq:tfb_MR},~\ref{eq:tvisc_MR}) and $\gg1$ for the stellar-mass case, so the rate supplied to the inner disc is a strongly time-variable function rather than a smooth quasi-steady decline, providing rapidly evolving launch conditions under which an angularly and velocity-stratified outflow may be established. By contrast, the long and well-separated timescales of a main-sequence--SMBH disruption ($\tfb\sim40$\,d, $\tfb\gg\tvisc$ for a realistic thick disc) yield a quasi-steady disc and a single outflow, perhaps of the form of a weakly evolving relativistic jet of the kind seen in Sw\,J1644+57 \citep{Berger2012,Eftekhari2018}.

To be more quantitative we compute, for each channel, the fallback rate $\Mdotfb(t)$ (Eq.~\ref{eq:Mdotfb}), the rate at which mass actually reaches the black hole $\dot M_{\rm acc}(t)$ and the rate at which mass is launched into outflows $\dot M_{\rm out}(t)$. We provide a derivation of how we define the accretion inflow and outflow rates in Appendix \ref{sec:TDE_timescales_outrate}. 

Figure~\ref{fig:mdot_compare} shows, for each channel, the no-wind inner rate $\dot M_0$ (Eq.~\ref{eq:Mdot0}) and the wind-partitioned $\Mdotacc$ and $\dot M_{\rm out}$ for $s=0.5$ and $1$. It is clear that the inflow $\dot M_0$ stays hugely super-Eddington for $\sim t_{\rm SE}$ (hundreds of days; for the WD--IMBH it is still $\sim10^{5}\,\Medd$ a day after disruption), so the engine easily outlives the prompt and X-ray phases. The accreted fraction is set by the number of radial decades over which the wind acts, $(r_{\rm I}/\Rcirc)^{s}$, and so depends strongly on the disc compactness: $\Rcirc/r_{\rm I}\approx9$, $900$, and $9\times10^{4}$ for the WD--IMBH, MS--IMBH, and sBH--MS channels respectively. Finally, the ejected fraction is extremely large for the extended-disc channels ($\gtrsim99.8\%$ (MS--IMBH) up to $\sim100\%$ (sBH--MS)) but more modest for the compact WD--IMBH disc, where $\sim66$--$89\%$ is ejected ($s=0.5$--$1$) and the black hole retains a genuinely super-Eddington accretion rate ($\sim0.06$--$0.17\,\Msun$ accreted). In every channel the bulk of the returned mass is nonetheless ejected, and these ejecta are naturally identified with an energetic, wide-angle, mildly relativistic outflow compatible with the radio. Launching even $\sim5\times10^{-4}\,\Msun$ at mildly relativistic speed supplies the inferred $E_K\sim10^{50}$\,erg, a tiny fraction of what is available. In this picture $t_{\rm SE}$ is simply the duration over which the wide-angle outflow is driven.
We note that the distribution of accretion disk X-ray luminosities of known TDEs (which presumably includes some super-Eddington sources) is strongly suppressed above $\sim 10^{44}$ erg/s, with an observed break in the luminosity function at this point \citep{Guolo2024, Grotova25, MummeryVV25}. This makes it very unlikely that the observed X-ray excess above the extrapolation of our synchrotron fits (at the level of $\sim 10^{48}$ erg/s) can be attributed to an accretion disk itself, as has been noted for similar X-ray excesses seen in other relativistic TDEs \citep[e.g.][]{Eftekhari2018}. 

\begin{figure*}
\centering
\includegraphics[width=\linewidth]{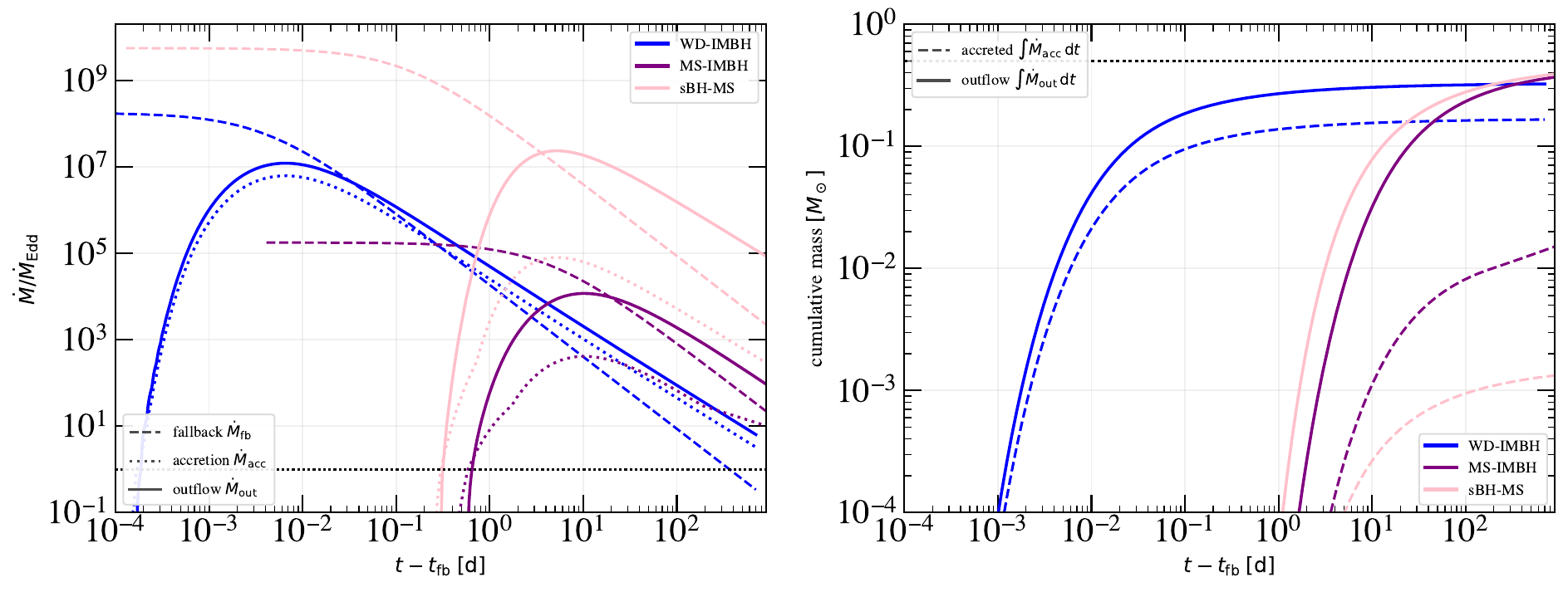}
\caption{Outflow rates of various TDE accretion flow at canonical wind index $s=1/2$, for the three viable channels (WD--IMBH, red; MS--IMBH, blue; sBH--MS, green). \textit{Left:} the fallback supply $\Mdotfb$ (dashed), the accretion rate onto the black hole $\Mdotacc$ (dotted), and the outflow rate $\dot M_{\rm out}=\dot M_0-\Mdotacc$ (solid), in Eddington units; the grey dotted line marks $\Medd$. \textit{Right:} the cumulative accreted mass $\int\Mdotacc\,{\rm d}t$ (dashed) and ejected mass $\int\dot M_{\rm out}\,{\rm d}t$ (solid); the grey dotted line marks $M_\star/2$. All curves assume $a=0.9$, $\alpha=0.1$, $H/R=0.3$, $\beta=1$, with the outflow region capped at the feeding radius, $R_{\rm out}=\min(R_{\rm sph},\Rcirc)$.  }
\label{fig:mdot_compare}
\end{figure*}

Taking the timescales together with the off-nuclear localisation, we conclude that all channels except an MS--SMBH disruption are consistent with the observed properties of GRB 250702B. The MS--SMBH channel is disfavoured both by the off-nuclear location \citep[][although wandering super massive black holes are known to power TDEs, e.g.,  \citealt{Guolo26}]{Levan2025} and by a fallback time ($\sim40$\,d) far longer than the observed prompt onset and precursor. We stress that, contrary to a na\"ive reading in which ``short timescales'' are taken to rule it out, the white-dwarf--IMBH channel is not excluded by the timescale analysis. Its short dynamical and viscous times reproduce the rapid variability, while its super-Eddington supply ($\sim300$\,d) sustains the engine through the prompt and X-ray phases. The one feature this channel does not produce naturally is the $\sim1$\,d EP precursor, since $\tfb\sim$\,minutes; reproducing it would require an additional ingredient such as the timing of stream self-intersections or a partial/successive disruption \citep{Eyles2026}. The precursor timing favours the channels whose fallback turns on at $\sim$\,hours to a day (sBH--MS and MS--IMBH) over the minutes-scale onset of the WD--IMBH, similar to the conclusions drawn by \citet{Beniamini2026} and \citet{Granot2026}.

\subsubsection{Relativistic jets from TDEs at different mass scales}

If the progenitor for GRB 250702B is a TDE around a stellar or intermediate mass black hole, it is interesting to consider why the relativistic jet from this system looks so different to the handful of observed relativistic jets from SMBH TDEs (Figure \ref{fig:grb_comp}). As recently shown by \citet{Decolle2026}, the sustained energy injection in relativistic TDE jets significantly modifies the dynamics and resulting radio emission compared to impulsive relativistic blastwaves produced in standard GRBs. The relativistic jets observed in SMBH TDEs, such as Swift J1644+57, are well described by a single, highly relativistic jet launched by a quasi-steady accretion flow \citep{Berger2012,Eftekhari2018,Rhodes2025}. In Swift J1644 (and other SMBH TDEs), fallback and viscous timescales are significantly longer (Table \ref{tab:TDE_timescales}). This naturally leads to a relatively stable accretion disk and a jet whose Lorentz factor and angular structure evolve only weakly with time. In contrast, for IMBH or stellar-mass disruption scenarios the fallback, circularisation, and accretion timescales are comparable (Table \ref{tab:TDE_timescales}), resulting in a dynamically evolving central engine in which the baryon loading and jet properties are expected to change significantly over time. This would likely change the observed structure of the jet. In the TDE scenario, sustained extremely super-Eddington accretion could drive the extended X-ray activity and a wider angle radio-producing outflow. The potential requirement in our radio observations of GRB 250702B for a wide angle outflow therefore points toward a progenitor in which the accretion flow is still in the process of forming and evolving, rather than the quasi-steady configuration expected in SMBH TDEs, further confirming the black hole in this system cannot be an SMBH.

The opening angle we infer for the late jet-break scenario of $\theta_j\gtrsim15$\,deg is consistent with the opening angle inferred for the TDE Sw J1644+57 of $\theta_j\approx20$\,deg \citep{Beniamini2023}, and broadly consistent with the idea that a larger mass black hole produces a less strongly collimated jet. An interesting discriminant between the TDE cases shown in Table \ref{tab:TDE_timescales} is the duration of super-Eddington accretion. Sw J1644+67 showed a sudden drop in X-ray flux at $\approx500$\,d post-TDE, interpreted as the jet-shut off at the time the accretion rate dropped below Eddington \citep{Zauderer2013,Eftekhari2018}. This is entirely consistent with the expected $\sim2$\,yr duration of super-Eddington accretion for a canonical solar mass star disrupted by a $10^6$\,M$_{\odot}$ black hole. The detection of a similar jet shut-off in GRB 250702B would provide a strong discriminant between the stellar mass or WD--IMBH scenarios (for which super-Eddington accretion should persist for $\sim300$\,d) or MS--IMBH/MS--sBH TDE channels (for which super-Eddington accretion persists for $10$\,yr); see the theoretical outflow rate profiles plotted in Figure \ref{fig:mdot_compare}. 

\section{Summary}\label{sec:summary}
In this work, we present extensive multi-epoch broadband radio observations of the first year of the radio afterglow produced by the transient GRB 250702B. Our conclusions are summarised as follows:

\begin{itemize}
    \item The radio spectra and evolution are consistent with an initially relativistic outflow where the spectrum transitions from $\nu_a < \nu_m < \nu_c$ to $\nu_m < \nu_a < \nu_c$ at $\sim50$\,d post-burst (observer frame). The evolution of the peak flux density and critical frequencies are broadly consistent with an adiabatic forward shock expanding into a stellar-wind or Bondi-like environment. Mild deviations from the simple self-similar forward shock solution may be due to a reverse shock component and/or jet structure, which future afterglow modelling efforts may aid in disentangling (Rohde et al. in prep). 
    \item Variability observed in the 1.25 and 3\,GHz light curves is consistent with interstellar scintillation, constraining a blastwave image size of $1.2\times10^{16}\lesssim R_{\perp} \lesssim 5\times10^{17}$\,cm at early times. 
    \item The lack of a jet break detected in the radio observations spanning 6--356\,d (observer frame) or the X-ray observations from $>0.5$\,d (observer frame) implies a jet half-opening angle of $\theta_j\lesssim2$\,deg or $\theta_j\gtrsim15$\,deg. Either a low-Lorentz factor, likely wide angle outflow or a narrow highly relativistic jet therefore remain viable options. 
    \item The observed evolution of the synchrotron emission constrains the density of the environment to be stratified, and likely consistent with a Bondi or stellar-wind like environment ($k=1.5$ to $2$) favouring both massive star and TDE-based progenitors.
    \item The timescale of evolution of the radio emission is more consistent with the classical GRB population, but the total luminosity, outflow energetics, and Lorentz factor is consistent with both the relativistic TDE population or GRB population, depending on the jet geometry. 
    \item We derive and show that the characteristic timescales for a TDE involving a stellar mass or intermediate black hole disrupting a main sequence or white dwarf star are compatible with the observed timescales of the prompt high energy emission observed and would naturally produce a wider angle jet, and the sustained highly super-Eddington accretion could supply the observed engine lifetime. An SMBH TDE is ruled out based on incompatibility with the observed timescales. 
\end{itemize}

The radio emission from GRB 250702B is rapidly fading, and will soon fade below detectable thresholds at all frequencies (with the current peak of the SED at $\approx$0.2\,mJy at $\approx$1\,GHz on 2026 June 06). Given the lack of detection of a jet-break to date, if the half-opening angle of the jet is $\gtrsim15$\,deg, we may still expect to detect a jet break, or jet shut-off when the accretion rate drops below super-Eddington, as has been observed for SMBH TDEs such as Sw J1644 \citep{Eftekhari2018}, and predicted to occur imminently for a WD-IMBH TDE scenario. Future high sensitivity observations at X-ray and radio frequencies may therefore constrain the geometry and energetics of the jet. Future efforts to incorporate the radio observations of GRB 250702B with multiwavelength observations in detailed afterglow modelling may provide crucial insight into the relative contributions of the forward and reverse shocks, as well as the jet opening angle and overall total energetics, which currently remain uncertain. 

\section*{Acknowledgments}
AJG thanks Wenbin Lu for helpful discussions. 
AJG is grateful for support from the Forrest Research Foundation. 
R. M. acknowledges partial support from the National
Science Foundation (grant number AST-2224255). KDA
and CTC acknowledge support provided by the NSF
through award SOSPA9-007 from the NRAO and award
AST-2307668. KDA gratefully acknowledges support
from the Alfred P. Sloan Foundation.
GCA thanks the Indian National Science Academy (INSA) for support under their Senior Scientist Programme.
MPT acknowledges financial support from the Severo Ochoa grant.
E.R-R. thanks the Heising-Simons Foundation, and NSF grants AST-1852393, AST-1911206, AST-2150255, AST-2206243 and 2447606 for support.
The National Radio Astronomy Observatory and Green Bank Observatory are facilities of the U.S. National Science Foundation operated under cooperative agreement by Associated Universities, Inc.
This paper makes use of the following ALMA data: 2023.1.01731.T. ALMA is a partnership of ESO (representing its member states), NSF (USA) and NINS (Japan), together with NRC (Canada), NSTC and ASIAA (Taiwan), and KASI (Republic of Korea), in cooperation with the Republic of Chile. The Joint ALMA Observatory is operated by ESO, AUI/NRAO and NAOJ.
We thank the staff of the GMRT that made these observations possible. GMRT is run by the National Centre for Radio Astrophysics of the Tata Institute of Fundamental Research.These observations were conducted under uGMRT ToO proposal 48\_059 (PI: D. Eappachen).

\section*{Data Availability}
All data are incorporated into the article (Table \ref{tab:all_fluxes}) and its online supplementary material.

\bibliographystyle{mnras}
\bibliography{example} 

\appendix

\section{The synchrotron electron index p}\label{sec:p_fits}

In the main text we fix $p=2.5$ in the synchrotron fits. This value is motivated by spectral fits to the radio data where the optically thin spectral slope is most well-constrained (at 57, 107, and 338\,d), as well as the spectral index between the mm and X-ray observations (shown in Figure \ref{fig:x-rayfits}). 

\begin{figure}
    \centering
    \includegraphics[width=\linewidth]{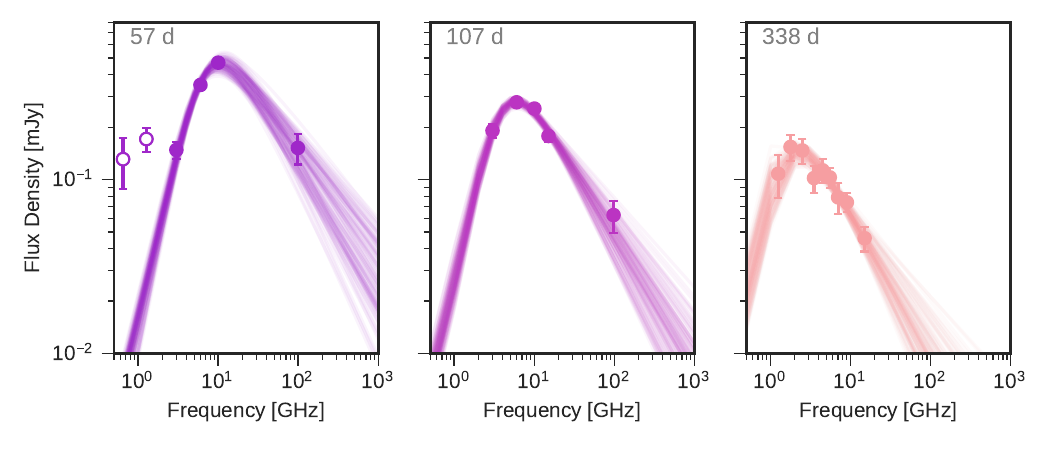}
        \includegraphics[width=0.5\linewidth]{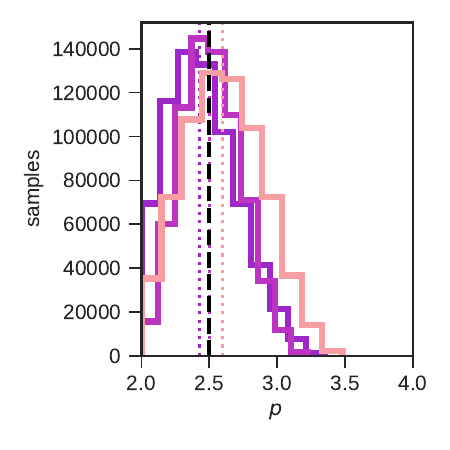}
    \caption{\textit{Top:} Synchrotron spectral fits with $p$ included as a fitting parameter for the radio data at the three epochs where the optically thin spectral slope is most well-constrained, allowing for a measurement of the synchrotron electron index, $p$. \textit{Bottom:} The posterior samples of $p$ for each of these 3 fits (solid colour lines). The median value of each histogram is plotted as a dotted line, and the total sample median, of $p=2.51$ is plotted as the black dashed line. }
    \label{fig:free_p_spec}
\end{figure}

The spectral fits with $p$ included as a fitting parameter (allowed to vary between $2<p<4$) for those epochs are shown in Figure \ref{fig:free_p_spec}. The posterior histograms for $p$ are also plotted, showing $p$ is well-constrained, with an average across the 3 fits of $p=2.51\pm0.1$. 

\section{Flux Density Extrapolation for epoch-by-epoch spectral fitting}\label{sec:flux_extrapolation}
\begin{figure}
    \centering
    \includegraphics[width=\linewidth]{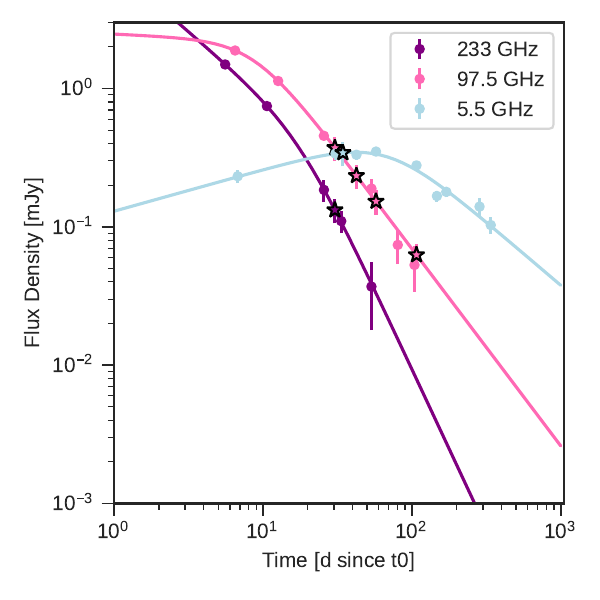}
    \caption{The observed (circles) 5.5, 97.5, and 233\,GHz light curves of GRB 250702B with simple broken power-law fits (solid lines) to each frequency. In order to obtain complete spectra for the spectral modelling and equipartition analysis, we extrapolated the flux density between epochs, to obtain approximate flux density measurements coincident with the majority of frequency points for each epoch. The extrapolated flux densities used in the spectral fitting are plotted as black outlined stars. }
    \label{fig:flux_extrapolation}
\end{figure}

The scheduling of simultaneous observations between facilities for this source was particularly challenging owing to weather constraints for the high frequency observations. Therefore, unfortunately, not all ALMA+VLA observations were able to be scheduled within a time frame that allows simultaneous spectral fitting without source temporal evolution between frequencies. In order to progress with the spectral fitting, we therefore carried out simple broken power-law fits of the 5.5, 97.5, and 233\,GHz light curves of the source and used this function to extrapolate between nearby epochs to obtain simultaneous flux density measurements of the source. For each extrapolated flux density measurement we include a conservative error of 20\% for the spectral fitting. 
A summary of the observed, fit, and extrapolated flux densities used in the spectral fitting is plotted in Figure \ref{fig:flux_extrapolation}. 

\section{Derivation of the relevant TDE Timescales}\label{sec:TDE_timescales}

A tidal disruption event has four distinct timescales, all of which control different elements of the evolution of the system. In this section we recap standard derivations \citep[e.g.,][]{rees_tidal_1988, Guillochon2013,Stone2019,gezari_tidal_2021} of all of the timescales. None of this section is new, we simply collate all derivations here for ease of discussion. 

To relate these timescales to observer times, we shall use the dimensionless normalisations
\begin{multline}
M_6 \equiv \frac{\Mbh}{10^6\,\Msun},\quad
m_\star \equiv \frac{M_\star}{\Msun},\quad
r_\star \equiv \frac{R_\star}{\Rsun},\quad
\alpha_{0.1} \equiv \frac{\alpha}{0.1},\quad
\theta_{0.3} \equiv \frac{H/R}{0.3},
\end{multline}
so that $\alpha_{0.1}\!=\!\theta_{0.3}\!=\!1$ corresponds to an $\alpha$
viscosity of $0.1$ \citep{ShakuraSunyaev1973} and a disk aspect ratio of $0.3$. This is deemed canonical for a (wildly) super-Eddington flow (which is likely to be the relevant limit for an IMBH TDE). 

The first is the fall-back timescale, or the time it takes the most-bound debris from a tidal disruption to return to pericentre on its (highly eccentric) orbit. The star itself approaches the black hole on
an almost-parabolic orbit and is disrupted when the tidal acceleration
exceeds its self-gravity, which occurs at the tidal radius
\begin{equation}
\rT \;=\; R_\star \left(\frac{\Mbh}{M_\star}\right)^{\!1/3}.
\label{eq:rT}
\end{equation}
At the moment of disruption different fluid elements of the star sit at
slightly different distances from the black hole.  In the
frozen-in energy approximation each element
retains the specific orbital energy it had at that instant, namely a
Keplerian energy evaluated at its own radius.  Linearising
$-G\Mbh/r$ across the star $r\!=\!\rT\!\pm\!R_\star$ gives an
energy spread
\begin{equation}
\Delta\varepsilon \;=\; \frac{G\Mbh\,R_\star}{\rT^2},
\end{equation}
so the most-bound debris has specific energy
$\varepsilon_{\min} = -\Delta\varepsilon$.  Its Keplerian semi-major axis
is therefore
\begin{equation}
a_{\min} \;=\; \frac{G\Mbh}{2|\varepsilon_{\min}|}
        \;=\; \frac{\rT^2}{2\,R_\star},
\end{equation}
and its orbital period (i.e. the time at which this most-bound debris first
returns to pericentre) is
\begin{equation}
\tfb \;=\; 2\pi\sqrt{\frac{a_{\min}^3}{G\Mbh}}
       \;=\; \frac{\pi}{\sqrt{2}}\,
             \sqrt{\frac{R_\star^3}{GM_\star}}\,
             \left(\frac{\Mbh}{M_\star}\right)^{\!1/2},
\label{eq:tfb_derived}
\end{equation}
having substituted Eq.~\eqref{eq:rT}.  In other words $\tfb$ is the stellar
dynamical time multiplied by $\sqrt{\Mbh/M_\star}$.  Inserting numbers,
\begin{equation}
\tfb \;\simeq\; 41\,{\rm d}\; M_6^{1/2}\, m_\star^{-1}\,
                                r_\star^{3/2}\;
\label{eq:tfb}
\end{equation}
 A flat distribution of energies
across the star then translates into the well-known $t^{-5/3}$ fallback
rate
\begin{equation}
\Mdotfb(t) \;=\; \frac{M_\star}{3\tfb}\,
                 \left(\frac{t}{\tfb}\right)^{-5/3}, \qquad t\ge\tfb,
\label{eq:Mdotfb}
\end{equation}
so $\tfb$ is both the time at which fallback begins and the
characteristic timescale on which the supply rate decays.  It depends
only on the orbital mechanics of the most-bound debris and nothing
about the disc enters $\tfb$.

The returning debris does not fall directly onto the black hole (it cannot do so, as that would violate angular momentum conservation).  Presumably, instead, it
shocks against itself, dissipates kinetic energy, and circularises on
an orbit whose specific angular momentum matches that of the original
orbit at pericentre.  The orbits are highly eccentric, and general-relativistic apsidal precession
advances the argument of periapsis each radial period.  For a geodesic with semi-major axis $a$ and eccentricity $e$, the advance per orbit is
\begin{equation}
\Delta\phi_{\rm ap}
\;=\;
\frac{6\pi\,\rg}{a\,(1-e^2)}
\;=\;
\frac{6\pi\,\rg}{R_p\,(1+e)},
\label{eq:dphi_ap}
\end{equation}
where $R_p = a(1-e)$ is the pericentre.  The most-bound TDE debris is
nearly parabolic ($e\to 1$), so
\begin{equation}
\Delta\phi_{\rm ap}
\;\simeq\;
\frac{3\pi\,\rg}{R_p}
\;=\;
\frac{3\pi\,\beta\,\rg}{\rT},
\label{eq:dphi_ecc}
\end{equation}
with impact parameter $\beta\!\equiv\!\rT/R_p$.  Inserting
Eq.~\eqref{eq:rT} and $\rg = G\Mbh/c^2$ gives the numerical scaling
\begin{equation}
\Delta\phi_{\rm ap}
\;\simeq\;
0.20\,{\rm rad}\;
\beta\, M_6^{2/3}\, m_\star^{1/3}\, r_\star^{-1}
\;\simeq\;
11^\circ\,
\beta\, M_6^{2/3}\, m_\star^{1/3}\, r_\star^{-1}.
\label{eq:dphi_norm}
\end{equation}
Because $\rT/\rg\!\propto\!\Mbh^{-2/3}$, heavier black holes produce
more apsidal advance at fixed $\beta$.

This offset misaligns the outbound and inbound legs of the debris stream, forcing a self-intersection shock that dissipates orbital energy and builds a disc near the circularisation radius, i.e. the radius whose specific angular momentum matches the pericentre
value
\begin{equation}
\Rcirc \;=\; \frac{2\,\rT}{\beta}
\label{eq:Rcirc}
\end{equation}
for a dimensionless penetration  parameter $\beta\!\equiv\!\rT/R_p$.  The first intersection of the most-bound debris occurs after
roughly one fallback orbit, but full circularisation generally takes
several shocks.  A useful geometrical estimate is that the stream must
accumulate an apsidal displacement of order one radian before its
successive windings intersect strongly.  This suggests
\begin{equation}
N_{\rm form}
\;\sim\;
\max\!\left[1,\frac{1}{\Delta\phi_{\rm ap}}\right],
\qquad
t_{\rm form} \;\sim\; N_{\rm form}\,\tfb .
\label{eq:Nform}
\end{equation}
Combining Eqs.~\eqref{eq:tfb}, \eqref{eq:dphi_norm}, and
\eqref{eq:Nform}, the weak-to-moderate-precession branch
($\Delta\phi_{\rm ap}\!\lesssim\!1$) is
\begin{equation}
t_{\rm form}
\;\sim\;
205\,{\rm d}\;
\beta^{-1}\,M_6^{-1/6}\,m_\star^{-4/3}\,r_\star^{5/2},
\qquad
\Delta\phi_{\rm ap}\lesssim 1,
\label{eq:tform}
\end{equation}
and it saturates at $t_{\rm form}\!\sim\!\tfb$ once
$\Delta\phi_{\rm ap}\!\gtrsim\!1$.  At canonical main-sequence
parameters, $\Delta\phi_{\rm ap}\!\simeq\!0.20$ and hence
$N_{\rm form}\!\sim\!5$.

Equation~\eqref{eq:Nform} should be regarded as an order-of-magnitude
geometrical estimate, not a calibrated circularisation law.  It reproduces the simulation trend that deeper encounters and more
massive black holes form discs faster, and its canonical value lies
within the commonly inferred range of one to ten most-bound periods
\citep{Bonnerot2016}.  However, kinematic calculations instead find a
fairly sharp transition between slow and fast circularisation near
$\Delta\phi_{\rm ap}\!\simeq\!0.2$, with no universal power-law relation
between $t_{\rm form}$ and $\Delta\phi_{\rm ap}$ \citep{Rossi2021}.  In the
weak-shock limit the energy dissipated per self-intersection scales
roughly as $\sin^2(\Delta\phi_{\rm ap}/2)$; demanding complete energetic
circularisation could therefore give the steeper
$N_{\rm form}\!\propto\!\Delta\phi_{\rm ap}^{-2}$ rather than the
geometrical $\Delta\phi_{\rm ap}^{-1}$ used here
\citep{LuBonnerot2020}.  Radiative cooling, stream thickness, black-hole
spin (nodal/Lense--Thirring precession), and repeated nozzle or
secondary shocks introduce further scatter.

The assumption that disk forms sweeps a lot of complex and poorly understood physics under the rug, but is consistent with the overwhelming observational evidence that accretion disks do form in TDEs \citep[e.g.,][]{vanvelzen2019,Mummery2024, Guolo2024}. 
The dynamical (orbital) time at the circularisation radius is
\begin{equation}
\;\torb \;=\; 2\pi\sqrt{\frac{\Rcirc^3}{G\Mbh}}\;
\label{eq:torb}
\end{equation}
For canonical parameters $M_6\!=\!1$, $m_\star\!=\!r_\star\!=\!\beta\!=\!1$
one finds $\Rcirc\!\approx\!94\,\rg$ and $\torb\!\approx\!0.33\,{\rm d}$.
Working through the algebra, $R_T \propto \Mbh^{1/3}$ exactly cancels the
$\Mbh^{-1/2}$ in the Kepler frequency, so $\torb$ is independent
of the black-hole mass and depends only on the stellar mean density and
the impact parameter:
\begin{equation}
\;\torb \;\simeq\; 0.33\,{\rm d}\;
       r_\star^{3/2}\, m_\star^{-1/2}\,\beta^{-3/2}\;
\label{eq:torb_norm}
\end{equation}
This is the shortest of the three timescales  and sets the
clock for everything inside the disc.  Unlike $\tfb$, it is independent of the black hole mass. 

Angular-momentum transport in an accretion disc operates on the
``viscous'' timescale
\begin{equation}
\tvisc \;\sim\; \frac{R^2}{\nu}, \qquad \nu \;\equiv\; \alpha c_s H,
\end{equation}
where $\alpha$ is the \citet{ShakuraSunyaev1973} viscosity parameter and $H$ is
the disc scale height.  Evaluating at $R = \Rcirc$, with the disc
isothermal so that $c_s = H \Omega_K$,
\begin{equation}
\tvisc \;=\; \frac{1}{\alpha\,\Omega_K(\Rcirc)} \left(\frac{R}{H}\right)^{\!2}
       \;=\; \frac{\torb}{2\pi\,\alpha\,(H/R)^2} ,
\label{eq:tvisc}
\end{equation}
The combination $\alpha (H/R)^2$ is what actually controls the disc's
spreading rate; the bare Shakura--Sunyaev $\alpha$ enters only through
this combination.  Inserting Eq.~\eqref{eq:torb_norm} and the canonical normalisation gives
\begin{equation}
\;\tvisc \;\simeq\; 5.8\,{\rm d}\;
        r_\star^{3/2}\, m_\star^{-1/2}\,\beta^{-3/2}\,
        \alpha_{0.1}^{-1}\, \theta_{0.3}^{-2}\;
\label{eq:tvisc_norm}
\end{equation}

A fourth, observationally useful timescale follows  from
Eq.~\eqref{eq:Mdotfb}, namely how long does the fallback rate stay above the
Eddington rate?  Setting $\Mdotfb(t_{\rm SE}) = \Medd$ with
\begin{equation}
\Medd \;=\; \frac{4\pi G \Mbh}{\kappa_{\rm es}\,\eta\,c}
\end{equation}
(where $\eta(a)$ is the Kerr radiative efficiency and
$\kappa_{\rm es}\!\simeq\!0.034\,{\rm m^2\,kg^{-1}}$ the electron-scattering
opacity) gives
\begin{equation}
t_{\rm SE} \;=\; \tfb \left(\frac{\Mdot_{\rm peak, fb}}{\Medd}\right)^{\!3/5},
\qquad
\Mdot_{\rm peak, fb} \;\equiv\; \frac{M_\star}{3\,\tfb}.
\label{eq:tSE_def}
\end{equation}
Note that this is a fallback-defined super-Eddington duration it states when the supply rate at $\Rcirc$ drops below $\Medd$, not when the
accretion rate onto the black hole drops below Eddington.  ``Viscous''
delay will shift the super-Eddington duration relative to
$t_{\rm SE}$, but for $\tvisc\!\ll\!\tfb$ the two might agree to factors of order unity (for $t_{\rm visc}\gtrsim t_{\rm fb}$ it is much longer).

The peak fallback rate scales as $\Mdot_{\rm peak, fb}\!\propto\!m_\star^{2}\,
\Mbh^{-3/2}\,r_\star^{-3/2}$, giving the famous result that lower-mass
black holes have longer super-Eddington phases.  Numerically,
\begin{equation}
\frac{\Mdot_{\rm peak, fb}}{\Medd}
\;\simeq\; 135\;\eta_{0.1}\,
m_\star^{2}\, M_6^{-3/2}\, r_\star^{-3/2},
\qquad
\eta_{0.1} \equiv \frac{\eta(a)}{0.1},
\end{equation}
and so
\begin{equation}
\;t_{\rm SE} \;\simeq\; 780\,{\rm d}\;
        M_6^{-2/5}\, m_\star^{1/5}\, r_\star^{3/5}\,
        \eta_{0.1}^{3/5}\;
\label{eq:tSE_norm}
\end{equation}
About two years for a canonical $10^6\Msun$ disruption of a Sun-like star, but shrinking to weeks for $M_6\!\gtrsim\!30$ and stretching to many years for intermediate-mass black holes.  

Dividing Eq.~\eqref{eq:tSE_norm} by Eq.~\eqref{eq:tfb} gives the
super-Eddington fallback duration in units of the fallback time
\begin{equation}
\;
\frac{t_{\rm SE}}{\tfb} \;\simeq\; 19\;
M_6^{-9/10}\, m_\star^{6/5}\, r_\star^{-9/10}\,
\eta_{0.1}^{3/5}\;.
\;
\label{eq:tSE_over_tfb}
\end{equation}
At canonical stellar parameters
($M_6\!=\!m_\star\!=\!r_\star\!=\!\eta_{0.1}\!=\!1$) the
super-Eddington-fallback phase lasts roughly an order of magnitude longer than
the fallback time itself. The  scaling
$\propto\Mbh^{-9/10}$  means that an IMBH will have super-Eddington fallback for $\sim 10^3$ fallback times. The $r_\star^{-9/10}$ scaling means that denser objects (white dwarfs) stay super-Eddington-(fallback) for longer (in units of fallback time).

The formulas above are written for canonical main-sequence parameters, but white-dwarf disruptions by intermediate-mass black holes follow trivially.
 Since $M_{\rm Hills}\propto R_\star^{3/2} M_\star^{-1/2}$,
the WD Hills mass (the maximum mass of a black hole which can disrupt an object outside of its horizon, \citealt{hills_possible_1975, Kesden12, Mummery24}) is smaller than the main-sequence value
$\sim\!10^{8}\Msun$ by a factor
$\sim (\Rsun/R_\oplus)^{3/2}
\!\sim\!10^{3}$, giving
$M_{\rm Hills,WD}\!\sim\!10^{5}\Msun$ at canonical parameters.  The
timescale numbers obviously drop,  the prompt phases shorten to
seconds--minutes, while the super-Eddington tail stretches to many
months.

We define the dimensionless WD/IMBH normalisations
\begin{equation}
m_{\rm WD} \equiv \frac{M_{\rm WD}}{\Msun},\quad
r_{\rm WD,\oplus} \equiv \frac{R_{\rm WD}}{R_\oplus},\quad
M_4 \equiv \frac{\Mbh}{10^4\,\Msun},
\end{equation}
with $R_\oplus/\Rsun\!\simeq\!9.16\!\times\!10^{-3}$ folding into every
$r_\star$ factor.  Re-running the four timescales with these
substitutions gives the following.  Eq.~\eqref{eq:tfb} becomes 
\begin{equation}
\;\tfb \;\simeq\; 5.1 \,{\rm min}\;
        M_4^{1/2}\, m_{\rm WD}^{-1}\, r_{\rm WD,\oplus}^{3/2}\;
\end{equation}
Five minutes for a canonical solar-mass, Earth-sized WD on a $10^4\Msun$
hole.

Equation \eqref{eq:torb_norm} has no $\Mbh$ dependence, so only the
stellar substitution applies:
\begin{equation}
\;\torb \;\simeq\; 25\,{\rm s}\;
        r_{\rm WD,\oplus}^{3/2}\, m_{\rm WD}^{-1/2}\,\beta^{-3/2}\;
\end{equation}
the disc's dynamical clock is now sub-minute.

The viscous time just follows $\torb$,
so the conversion is the same:
\begin{equation}
\;\tvisc \;\simeq\; 7.3\,{\rm min}\;
        r_{\rm WD,\oplus}^{3/2}\, m_{\rm WD}^{-1/2}\,\beta^{-3/2}\,
        \alpha^{-1}_{0.1}\,\theta_{0.3}^{-2}\;
\end{equation}
Notably $\tvisc/\tfb \sim {\cal O}(1)$ at the WD canonical point -- the
hierarchy $\tvisc\!\ll\!\tfb$ that held for thick-disc main-sequence TDEs no longer
applies, and ``viscous'' smearing of the fallback peak becomes a first-order effect even for thick discs.

For Eq.~\eqref{eq:tSE_norm} the conversion factors give 
\begin{equation}
\;t_{\rm SE} \;\simeq\; 295\,{\rm d}\;
        M_4^{-2/5}\, m_{\rm WD}^{1/5}\, r_{\rm WD,\oplus}^{3/5}\,
        \eta_{0.1}^{3/5}\;
\end{equation}
Ten months above Eddington for the canonical WD--IMBH disruption,
because the peak fallback rate
$\Mdot_{\rm peak}/\Medd\!\simeq\!1.5\!\times\!10^{8}\,\eta_{0.1}\,
m_{\rm WD}^{2}\,M_4^{-3/2}\,r_{\rm WD,\oplus}^{-3/2}$
beats Eddington by eight orders of magnitude at peak.  The ratio
$t_{\rm SE}/\tfb \!\simeq\!  8\!\times\!10^{4}$ in this regime --
several orders of magnitude larger than for main-sequence TDEs.

Finally, the formulas above treat $m_{\rm WD}$ and $r_{\rm WD,\oplus}$ as
independent knobs, but real white dwarfs obey a non-relativistic
degeneracy-supported mass--radius relation
\begin{equation}
r_{\rm WD,\oplus} \;\simeq\; m_{\rm WD}^{-1/3},
\label{eq:WD_MR}
\end{equation}
calibrated at $m_{\rm WD}\!=\!1$ (more massive WDs are denser and
smaller, with the relation changing as one approaches the
Chandrasekhar mass $1.4\Msun$ and relativistic corrections kick in).
Substituting Eq.~\eqref{eq:WD_MR} into the relevant timescales:
\begin{align}
\tfb        &\;\simeq\; 5.1\,{\rm min}\;
              M_4^{1/2}\, m_{\rm WD}^{-3/2}, \label{eq:tfb_MR}\\
\torb       &\;\simeq\; 25\,{\rm s}\;
              m_{\rm WD}^{-1}\,\beta^{-3/2}, \label{eq:torb_MR}\\
\tvisc      &\;\simeq\; 7.3\,{\rm min}\;
              m_{\rm WD}^{-1}\,\beta^{-3/2}\,
              \alpha_{-1}^{-1}\,\theta_{-0.5}^{-2}, \label{eq:tvisc_MR}\\
t_{\rm SE}  &\;\simeq\; 295\,{\rm d}\;
              M_4^{-2/5}\, \eta_{0.1}^{3/5}. \label{eq:tSE_MR}
\end{align}
The most interesting result is Eq.~\eqref{eq:tSE_MR}, the
super-Eddington duration is independent of the WD mass.

\subsection{Accretion rate and outflow rate}\label{sec:TDE_timescales_outrate}

Material returning to $\Rcirc$ is not delivered to the black hole instantaneously (it cannot be as this would not conserve angular momentum), instead it diffuses inward on the ``viscous'' timescale. Assuming that the disc surface density obeys a linear diffusion equation (perhaps reasonable in the scale-free super Eddington limit), the accretion rate in the absence of mass loss is the convolution of the fallback injection history at $\Rcirc$ with the disc's impulse response (Green's function) $G_{\Mdot}$ \citep{Mummery2023},
\begin{equation}\label{eq:Mdot0}
\dot M_{0}(t) = \frac{M_\star}{3\tfb}\int_{\tfb}^{t}\left(\frac{t'}{\tfb}\right)^{-5/3} G_{\Mdot}\!\left(r_{\rm I},t\,\middle|\,t',\Rcirc\right){\rm d}
 t',
\end{equation}
with $G_{\Mdot}$ normalised to conserve mass, $\int_0^\infty G_{\Mdot}\,{\rm d}\tau = 1$. This $\delta$-function source assumes prompt circularisation of the returning debris into a disc at $\Rcirc$, which is best justified when $\torb\ll\tfb$. This holds comfortably for the IMBH channels but is only marginal for the sBH--MS case, where $\torb\gtrsim\tfb$ (Table~\ref{tab:TDE_timescales}) and the debris may cannot fully circularise before subsequent material returns. Indeed, the assumed ordering $\torb<\tfb$ only holds for $\Mbh\gtrsim64\,M_\star$ (the point at which the most-bound semi-major axis $a_{\min}=\rT^2/2R_\star$ exceeds $\Rcirc=2\rT$). 
The sBH--MS timescales should accordingly be read as order-of-magnitude indicators only. 

At the Eddington ratios reached here, however, $\dot M_0$ is not the accretion rate that actually reaches the black hole, because within the super-Eddington region the disc drives strong outflows. The formal ``spherization'' radius, $R_{\rm sph}\simeq\dot m_0\,r_{\rm I}$ \citep{ShakuraSunyaev1973} with $\dot m_0\equiv\dot M_0/\Medd$, lies far outside the disc itself. For the WD--IMBH at peak, $\dot m\sim10^{8}$, versus a circularisation (feeding) radius $\Rcirc\sim20\,\rg$ so the entire disc, out to where it is fed, is super-Eddington and outflow-driving. 
Adopting the standard self-similar prescription $\dot M(R)\propto R^{s}$ over the disc extent \citep{ShakuraSunyaev1973}, the fraction reaching the ISCO is $(r_{\rm I}/R_{\rm out})^{s}$, so that
\begin{align}\label{eq:wind}
\Mdotacc(t)&=\dot M_0(t)\left[\min\!\left(\frac{\Rcirc}{r_{\rm I}},\,\max(1,\dot m_0)\right)\right]^{-s},\\
\dot M_{\rm out}(t) &=\dot M_0-\Mdotacc.
\end{align}
When the inflow is only mildly super-Eddington ($1<\dot m_0<\Rcirc/r_{\rm I}$) the outflow region is set by $R_{\rm sph}$ and $\Mdotacc=\Medd\,\dot m_0^{1-s}$, the familiar result that enforces the Eddington limit onto the hole $\Medd$ (for $s=1$). But at the extreme rates here ($\dot m_0\gg\Rcirc/r_{\rm I}$) the cap at the feeding radius takes over, the accreted fraction saturates at the constant $(r_{\rm I}/\Rcirc)^{s}$, so the hole accretes a fixed fraction of the (still super-Eddington) inflow, with the trapped radiation advected inward rather than being held at $\Medd$. 
This is a post-hoc partition, not a self-consistent wind solution \citep{Mummery26} it ignores the angular momentum the wind removes (which would itself modify $\dot M_0$) and depends on $s$, on the order-unity factor in $R_{\rm sph}$, and on the disc outer radius (taken as $\Rcirc$, in reality the outward viscous spreading would push it larger, increasing the ejected fraction).

\newpage
\onecolumn

\section{Flux Density Measurements}
All new flux density measurements reported in this work are shown in Table \ref{tab:all_fluxes}.
\begin{longtable}{lllll}
\hline
Time (UTC) & Time (d) & Instrument & Frequency (GHz) & Flux Density (mJy) \\
\hline
2025-07-08 03:40:31& $5.573$ & ALMA & $233$ & $1.5 \pm 0.078$ \\
2025-07-08 05:29:18 & $5.648$ & VLA & $10$ & $0.49 \pm 0.025$ \\
2025-07-09 01:54:10 & $6.499$ & ALMA & $97$ & $1.9 \pm 0.097$ \\
2025-07-09 08:22:53 & $6.769$ & VLA & $3$ & $0.069 \pm 0.022$ \\
2025-07-09 08:22:53 & $6.769$ & VLA & $4.5$ & $0.18 \pm 0.028$ \\
2025-07-09 08:22:53 & $6.769$ & VLA & $5.5$ & $0.23 \pm 0.025$ \\
2025-07-09 08:22:53 & $6.769$ & VLA & $6.5$ & $0.29 \pm 0.032$ \\
2025-07-09 08:22:53 & $6.769$ & VLA & $7.5$ & $0.41 \pm 0.031$ \\
2025-07-09 08:22:53 & $6.769$ & VLA & $8.5$ & $0.44 \pm 0.03$ \\
2025-07-09 08:22:53 & $6.769$ & VLA & $9.3$ & $0.49 \pm 0.03$ \\
2025-07-09 08:22:53 & $6.769$ & VLA & $10$ & $0.57 \pm 0.035$ \\
2025-07-09 08:22:53 & $6.769$ & VLA & $11$ & $0.69 \pm 0.049$ \\
2025-07-09 08:22:53 & $6.769$ & VLA & $13$ & $0.71 \pm 0.054$ \\
2025-07-09 08:22:53 & $6.769$ & VLA & $14$ & $0.8 \pm 0.044$ \\
2025-07-09 08:22:53 & $6.769$ & VLA & $16$ & $0.89 \pm 0.049$ \\
2025-07-09 08:22:53 & $6.769$ & VLA & $17$ & $1 \pm 0.057$ \\
2025-07-09 08:22:53 & $6.769$ & VLA & $19$ & $1.3 \pm 0.072$ \\
2025-07-09 08:22:53 & $6.769$ & VLA & $21$ & $1.4 \pm 0.089$ \\
2025-07-09 08:22:53 & $6.769$ & VLA & $23$ & $1.6 \pm 0.094$ \\
2025-07-09 08:22:53 & $6.769$ & VLA & $25$ & $1.5 \pm 0.09$ \\
2025-07-09 08:22:53 & $6.769$ & VLA & $31$ & $1.7 \pm 0.095$ \\
2025-07-09 08:22:53 & $6.769$ & VLA & $32$ & $1.7 \pm 0.097$ \\
2025-07-09 08:22:53 & $6.769$ & VLA & $34$ & $1.8 \pm 0.097$ \\
2025-07-09 08:22:53 & $6.769$ & VLA & $36$ & $1.8 \pm 0.099$ \\
2025-07-12 19:22:59 & $9.419$ & GMRT & $1.3$ & $0.082 \pm 0.017$ \\
2025-07-13 04:59:55 & $10.628$ & ALMA & $233$ & $0.75 \pm 0.044$ \\
2025-07-15 05:22:29& $12.643$ & ALMA & $97$ & $1.1 \pm 0.061$ \\
2025-07-15 09:31:16 & $12.816$ & VLA & $3$ & $0.076 \pm 0.017$ \\
2025-07-15 09:31:16 & $12.816$ & VLA & $5$ & $0.24 \pm 0.02$ \\
2025-07-15 09:31:16 & $12.816$ & VLA & $6.5$ & $0.37 \pm 0.029$ \\
2025-07-15 09:31:16 & $12.816$ & VLA & $7.5$ & $0.41 \pm 0.029$ \\
2025-07-15 09:31:16 & $12.816$ & VLA & $8.5$ & $0.43 \pm 0.032$ \\
2025-07-15 09:31:16 & $12.816$ & VLA & $9.3$ & $0.5 \pm 0.034$ \\
2025-07-15 09:31:16 & $12.816$ & VLA & $10$ & $0.59 \pm 0.039$ \\
2025-07-15 09:31:16 & $12.816$ & VLA & $11$ & $0.7 \pm 0.045$ \\
2025-07-15 09:31:16 & $12.816$ & VLA & $13$ & $0.74 \pm 0.043$ \\
2025-07-15 09:31:16 & $12.816$ & VLA & $14$ & $0.83 \pm 0.046$ \\
2025-07-15 09:31:16 & $12.816$ & VLA & $16$ & $0.79 \pm 0.048$ \\
2025-07-15 09:31:16 & $12.816$ & VLA & $17$ & $0.86 \pm 0.055$ \\
2025-07-15 09:31:16 & $12.816$ & VLA & $19$ & $0.91 \pm 0.056$ \\
2025-07-15 09:31:16 & $12.816$ & VLA & $21$ & $0.84 \pm 0.073$ \\
2025-07-15 09:31:16 & $12.816$ & VLA & $23$ & $1.1 \pm 0.088$ \\
2025-07-15 09:31:16 & $12.816$ & VLA & $25$ & $1 \pm 0.077$ \\
2025-07-15 09:31:16 & $12.816$ & VLA & $31$ & $1.1 \pm 0.081$ \\
2025-07-15 09:31:16 & $12.816$ & VLA & $32$ & $1 \pm 0.073$ \\
2025-07-15 09:31:16 & $12.816$ & VLA & $34$ & $1 \pm 0.073$ \\
2025-07-15 09:31:16 & $12.816$ & VLA & $36$ & $0.95 \pm 0.081$ \\
2025-07-28 05:31:10 & $25.649$ & ALMA & $97$ & $0.46 \pm 0.034$ \\
2025-07-28 06:03:36 & $25.672$ & ALMA & $233$ & $0.18 \pm 0.034$ \\
2025-07-30 17:37:20 & $27.419$ & GMRT & $0.65$ & $0.12 \pm 0.042$ \\
2025-08-02 06:22:14 & $30.685$ & VLA & $3$ & $0.18 \pm 0.026$ \\
2025-08-02 06:22:14 & $30.685$ & VLA & $4.5$ & $0.22 \pm 0.033$ \\
2025-08-02 06:22:14 & $30.685$ & VLA & $5.5$ & $0.34 \pm 0.032$ \\
2025-08-02 06:22:14 & $30.685$ & VLA & $6.5$ & $0.34 \pm 0.03$ \\
2025-08-02 06:22:14 & $30.685$ & VLA & $7.5$ & $0.41 \pm 0.029$ \\
2025-08-02 06:22:14 & $30.685$ & VLA & $8.5$ & $0.48 \pm 0.032$ \\
2025-08-02 06:22:14 & $30.685$ & VLA & $9.3$ & $0.43 \pm 0.03$ \\
2025-08-02 06:22:14 & $30.685$ & VLA & $10$ & $0.47 \pm 0.035$ \\
2025-08-02 06:22:14 & $30.685$ & VLA & $11$ & $0.53 \pm 0.046$ \\
2025-08-02 06:22:14 & $30.685$ & VLA & $13$ & $0.58 \pm 0.05$ \\
2025-08-02 06:22:14 & $30.685$ & VLA & $14$ & $0.55 \pm 0.033$ \\
2025-08-02 06:22:14 & $30.685$ & VLA & $16$ & $0.58 \pm 0.038$ \\
2025-08-02 06:22:14 & $30.685$ & VLA & $17$ & $0.65 \pm 0.043$ \\
2025-08-05 03:29:33 & $33.565$ & ALMA & $233$ & $0.11 \pm 0.02$ \\
2025-08-05 04:18:06 & $33.599$ & ALMA & $97$ & $0.34 \pm 0.029$ \\
2025-08-06 05:31:18 & $34.649$ & VLA & $19$ & $0.7 \pm 0.052$ \\
2025-08-06 05:31:18 & $34.649$ & VLA & $21$ & $0.76 \pm 0.069$ \\
2025-08-06 05:31:18 & $34.649$ & VLA & $23$ & $0.74 \pm 0.07$ \\
2025-08-06 05:31:18 & $34.649$ & VLA & $25$ & $0.68 \pm 0.056$ \\
2025-08-06 05:31:18 & $34.649$ & VLA & $31$ & $0.66 \pm 0.077$ \\
2025-08-06 05:31:18 & $34.649$ & VLA & $32$ & $0.61 \pm 0.064$ \\
2025-08-06 05:31:18 & $34.649$ & VLA & $34$ & $0.62 \pm 0.056$ \\
2025-08-06 05:31:18 & $34.649$ & VLA & $36$ & $0.62 \pm 0.065$ \\
2025-08-14 00:47:09 & $42.452$ & VLA & $1.5$ & $0.28 \pm 0.08$ \\
2025-08-14 00:47:09 & $42.452$ & VLA & $2.5$ & $0.21 \pm 0.037$ \\
2025-08-14 00:47:09 & $42.452$ & VLA & $3.5$ & $0.23 \pm 0.027$ \\
2025-08-14 00:47:09 & $42.452$ & VLA & $4.5$ & $0.27 \pm 0.029$ \\
2025-08-14 00:47:09 & $42.452$ & VLA & $5.5$ & $0.33 \pm 0.028$ \\
2025-08-14 00:47:09 & $42.452$ & VLA & $6.5$ & $0.34 \pm 0.029$ \\
2025-08-14 00:47:09 & $42.452$ & VLA & $7.5$ & $0.42 \pm 0.03$ \\
2025-08-14 00:47:09 & $42.452$ & VLA & $8.5$ & $0.41 \pm 0.027$ \\
2025-08-14 00:47:09 & $42.452$ & VLA & $9.5$ & $0.45 \pm 0.029$ \\
2025-08-14 00:47:09 & $42.452$ & VLA & $10$ & $0.47 \pm 0.035$ \\
2025-08-14 00:47:09 & $42.452$ & VLA & $12$ & $0.55 \pm 0.072$ \\
2025-08-14 00:47:09 & $42.452$ & VLA & $13$ & $0.59 \pm 0.044$ \\
2025-08-14 00:47:09 & $42.452$ & VLA & $14$ & $0.59 \pm 0.036$ \\
2025-08-14 00:47:09 & $42.452$ & VLA & $16$ & $0.65 \pm 0.039$ \\
2025-08-14 00:47:09 & $42.452$ & VLA & $17$ & $0.61 \pm 0.041$ \\
2025-08-25 02:18:49 & $53.516$ & ALMA & $233$ & $0.037 \pm 0.019$ \\
2025-08-25 03:00:39 & $53.545$ & ALMA & $99$ & $0.19 \pm 0.031$ \\
2025-08-26 18:50:14 & $55.204$ & GMRT & $1.3$ & $0.17 \pm 0.027$ \\
2025-08-28 16:40:32 & $57.114$ & GMRT & $0.65$ & $0.13 \pm 0.043$ \\
2025-08-29 00:49:48 & $57.454$ & VLA & $3$ & $0.15 \pm 0.017$ \\
2025-08-29 00:49:48 & $57.454$ & VLA & $6$ & $0.35 \pm 0.021$ \\
2025-08-29 00:49:48 & $57.454$ & VLA & $10$ & $0.47 \pm 0.027$ \\
2025-09-09 13:20:34 & $68.975$ & GMRT & $1.3$ & $0.23 \pm 0.074$ \\
2025-09-11 14:43:45 & $71.033$ & GMRT & $0.65$ & $0.12 \pm 0.044$ \\
2025-09-21 00:47:31 & $80.452$ & ALMA & $98$ & $0.074 \pm 0.02$ \\
2025-10-14 18:12:50 & $104.178$ & ALMA & $97$ & $0.053 \pm 0.019$ \\
2025-10-18 00:56:45 & $107.459$ & VLA & $3$ & $0.19 \pm 0.018$ \\
2025-10-18 00:56:45 & $107.459$ & VLA & $6$ & $0.28 \pm 0.016$ \\
2025-10-18 00:56:45 & $107.459$ & VLA & $10$ & $0.26 \pm 0.016$ \\
2025-10-18 00:56:45 & $107.459$ & VLA & $15$ & $0.18 \pm 0.013$ \\
2025-10-29 01:59:53 & $118.503$ & VLA & $15$ & $0.17 \pm 0.013$ \\
2025-10-29 01:59:53 & $118.503$ & VLA & $22$ & $0.12 \pm 0.011$ \\
2025-10-29 01:59:53 & $118.503$ & VLA & $33$ & $0.1 \pm 0.015$ \\
2025-11-26 22:01:12 & $147.337$ & VLA & $3$ & $0.14 \pm 0.023$ \\
2025-11-26 22:01:12 & $147.337$ & VLA & $6$ & $0.17 \pm 0.015$ \\
2025-11-26 22:01:12 & $147.337$ & VLA & $9$ & $0.17 \pm 0.015$ \\
2025-11-26 22:01:12 & $147.337$ & VLA & $11$ & $0.15 \pm 0.013$ \\
2025-12-19 21:13:53 & $170.304$ & VLA & $3$ & $0.19 \pm 0.015$ \\
2025-12-19 21:13:53 & $170.304$ & VLA & $3$ & $0.19 \pm 0.016$ \\
2025-12-19 21:13:53 & $170.304$ & VLA & $6$ & $0.18 \pm 0.016$ \\
2025-12-19 21:13:53 & $170.304$ & VLA & $9$ & $0.15 \pm 0.012$ \\
2025-12-19 21:13:53 & $170.304$ & VLA & $11$ & $0.16 \pm 0.015$ \\
2025-12-19 21:13:53 & $170.304$ & VLA & $14$ & $0.18 \pm 0.013$ \\
2025-12-19 21:13:53 & $170.304$ & VLA & $16$ & $0.18 \pm 0.014$ \\
2026-01-10 18:29:19 & $192.190$ & VLA & $15$ & $0.14 \pm 0.012$ \\
2026-01-10 18:29:19 & $192.190$ & VLA & $22$ & $0.11 \pm 0.012$ \\
2026-01-10 18:29:19 & $192.190$ & VLA & $33$ & $<0.072$ \\
2026-04-13 11:47:04 & $284.910$ & VLA & $1.5$ & $0.15 \pm 0.021$ \\
2026-04-13 11:47:04 & $284.910$ & VLA & $2.2$ & $0.19 \pm 0.06$ \\
2026-04-13 11:47:04 & $284.910$ & VLA & $2.8$ & $0.16 \pm 0.027$ \\
2026-04-13 11:47:04 & $284.910$ & VLA & $3.2$ & $0.16 \pm 0.019$ \\
2026-04-13 11:47:04 & $284.910$ & VLA & $3.8$ & $0.15 \pm 0.025$ \\
2026-04-13 11:47:04 & $284.910$ & VLA & $4.5$ & $0.17 \pm 0.025$ \\
2026-04-13 11:47:04 & $284.910$ & VLA & $5.5$ & $0.14 \pm 0.02$ \\
2026-04-13 11:47:04 & $284.910$ & VLA & $7.3$ & $0.12 \pm 0.014$ \\
2026-04-13 11:47:04 & $284.910$ & VLA & $9$ & $0.097 \pm 0.011$ \\
2026-04-13 11:47:04 & $284.910$ & VLA & $11$ & $0.072 \pm 0.013$ \\
2026-05-15 09:48:17 & $316.828$ & VLA & $15$ & $0.042 \pm 0.01$ \\
2026-06-06 08:09:09 & $338.759$ & VLA & $1.2$ & $0.11 \pm 0.03$ \\
2026-06-06 08:09:09 & $338.759$ & VLA & $1.8$ & $0.15 \pm 0.026$ \\
2026-06-06 08:09:09 & $338.759$ & VLA & $2.5$ & $0.15 \pm 0.024$ \\
2026-06-06 08:09:09 & $338.759$ & VLA & $3.5$ & $0.1 \pm 0.018$ \\
2026-06-06 08:09:09 & $338.759$ & VLA & $4.5$ & $0.11 \pm 0.018$ \\
2026-06-06 08:09:09 & $338.759$ & VLA & $5.5$ & $0.1 \pm 0.014$ \\
2026-06-06 08:09:09 & $338.759$ & VLA & $7$ & $0.079 \pm 0.016$ \\
2026-06-07 07:40:14 & $339.739$ & VLA & $9$ & $0.074 \pm 0.0093$ \\
2026-06-07 07:40:14 & $339.739$ & VLA & $15$ & $0.046 \pm 0.0074$ \\
2026-06-23 04:44:19  & $355.617$ & VLA &  $22$ & $0.029\pm0.008$ \\
2026-06-23 06:14:03  & $355.679$ & VLA &  $33$ & $0.023\pm0.008$ \\
\hline
\caption{Radio flux density measurements of GRB 250702B presented in this work. Time in days is measured since MJD $60858.58061$.}
\label{tab:all_fluxes}
\end{longtable}

\bsp	
\label{lastpage}
\end{document}